\documentclass[sn-mathphys,Numbered]{sn-jnl}

\usepackage{graphicx}%
\usepackage{multirow}%
\usepackage{amsmath,amssymb,amsfonts}%
\usepackage{amsthm}%
\usepackage{mathrsfs}%
\usepackage[title]{appendix}%
\usepackage{xcolor}%
\usepackage{textcomp}%
\usepackage{manyfoot}%
\usepackage{booktabs}%
\usepackage{algorithm}%
\usepackage{algorithmicx}%
\usepackage{algpseudocode}%
\usepackage{listings}%

\theoremstyle{thmstyleone}%
\theoremstyle{thmstyletwo}%
\theoremstyle{thmstylethree}%

\begin{document}

\title[Vortex Depinning in a Two-dimensional Superfluid]{Vortex depinning in a two-dimensional superfluid}


\author*[1]{\fnm{I-Kang} \sur{Liu}}\email{i-kang.liu1@newcastle.ac.uk}

\author[1]{\fnm{Srivatsa B.} \sur{Prasad}}\email{srivatsa.badariprasad@newcastle.ac.uk}

\author[1]{\fnm{Andrew W.} \sur{Baggaley}}\email{andrew.baggaley@newcastle.ac.uk}

\author[1]{\fnm{Carlo F.} \sur{Barenghi}}\email{carlo.barenghi@newcastle.ac.uk}

\author[1]{\fnm{Toby S.} \sur{Wood}}\email{toby.wood@newcastle.ac.uk}

\affil*[1]{\orgdiv{School of Mathematics, Statistics and Physics}, \orgname{Newcastle University}, \orgaddress{ \city{Newcastle upon Tyne}, \postcode{NE1 7RU}, \state{Tyne and Wear}, \country{United Kingdom}}}




\abstract{

We employ the Gross--Pitaevskii theory to model a quantized vortex depinning from a small obstacle in a two-dimensional superfluid due to an imposed background superfluid flow. We find that, when the flow's velocity exceeds a critical value, the vortex drifts orthogonally to the flow before subsequently moving parallel to it away from the pinning site. The motion of the vortex around the pinning site is also accompanied by an emission of a spiral-shaped sound pulse. Through simulations, we present a phase diagram of the critical flow velocity for vortex depinning together with an empirical formula that illustrates how the critical velocity increases with the height and width of the pinning site. By employing a variety of choices of initial and boundary conditions, we are able to obtain lower and upper bounds on the critical velocity and demonstrate the robustness of these results.

}





\maketitle

\section{Introduction}\label{sec:introduction}



The interaction between topological defects and their environment is responsible for a host of fascinating phenomena in physical and biological systems. One of the most important examples thereof is the pinning of topological defects, e.g. rotating waves in cardiac muscles~\cite{davidenko1992stationary,pumir1999unpinning,Takagi1991}, vortices in active matter~\cite{PhysRevLett.93.168303} and nematic defects in liquid crystals~\cite{PhysRevLett.112.197801,nayek2012tailoring}. In quantum fluids, such as superfluids and superconductors, the nucleation and motion of quantized vortices play crucial roles. For instance, in superfluid He-II, the presence of vortices can lead to dissipation of the superflow~\cite{donnelly1991quantized}. Additionally, vortices can be readily pinned to obstacles with a length scale comparable to the superfluid healing length~\cite{Tsubota1993}, such as bumps in a superfluid container and defects in superconductors. Furthermore, vortex pinning results in a correction to the Berezinskii--Kosterlitz--Thouless transition in thin-film He-II~\cite{Hegde1980,PhysRevB.21.1806,PhysRevB.35.4633}, holding magnetic flux in type-II superconductors~\cite{PhysRevB.48.13060,Blatter1994,kwok2016vortices}, an increase in critical counterflow velocity~\cite{PhysRevLett.112.197801}. Vortex pinning can also be implemented in order to manipulate atomtronic devices~\cite{PhysRevLett.111.235301,Amico:2020eej,Bland2022}.

Quantum fluids are also thought to be present in cosmological and astrophysical systems such as dark matter and neutron stars. In the ultra-light dark matter model~\cite{Schive2014, Marsh2015, Marsh2016, Ferreira}, vortices are found to be unstable in the central region of dark matter halos~\cite{Rindler-Daller2012,Dmitriev2021,Schobesberger2021} but are associated with the granule size with a turbulence-like characteristics in the outer regions~\cite{Liu2023}. In the interior of a neutron star, both neutrons and protons can be in the superfluid phase~\cite{Pethick2017}. The neutron fluid typically contains of order $10^{18}$ vortices, which can pin to the nuclear lattice in the star's outer crust~\cite{Jones2001, Donati2004} and to magnetic flux tubes in the star's core~\cite{Drummond2017a, Drummond2018}. In this system, vortex pinning prevents the superfluid from spinning down at the same rate as the crust, thereby creating a rotational lag. It is believed that, when this lag reaches a critical value, the vortices depin and transfer angular momentum from the superfluid to the crust, resulting in a sudden increase in the observed rotational frequency known as a glitch. The exact process by which a glitch occurs is not fully understood but may involve an avalanche of millions of vortices that depin and thereby trigger further depinning~\cite{Lonnborn2019}. Given the complexity of the mechanism underlying these spindown glitches, it is pertinent to first study the dynamics of a small number of vortices and thereby understand the conditions under which they depin, which is the aim of the present work.

From the perspective of a pinned vortex, the rotational lag in the neutron star crust manifests as an ambient superflow, which exerts a lateral Magnus force on the vortex~\citep{Sonin1997}. Above a critical superfluid velocity, this force causes the vortex to depin, after which it moves with essentially the same velocity as the ambient flow. This critical velocity was first studied numerically by Schwarz~\cite{Schwarz1981} using the vortex filament method (in which a vortex is modelled as a one-dimensional line); in this model, the pinning site was a hemispherical `bump' on the boundary of the superfluid. Subsequently, \citet{Tsubota1993} demonstrated that multiple vortices can be trapped on the same bump. More recently, \citet{Stockdale2021} investigated a related problem --- the scattering of a superfluid vortex by an obstacle. For an obstacle of large width, they found that the critical velocity for pinning could be predicted using the equation of motion for a vortex under the small-displacement approximation~\cite{Groszek2018}.

\begin{figure}
\begin{center}    \includegraphics[width=0.6\textwidth]{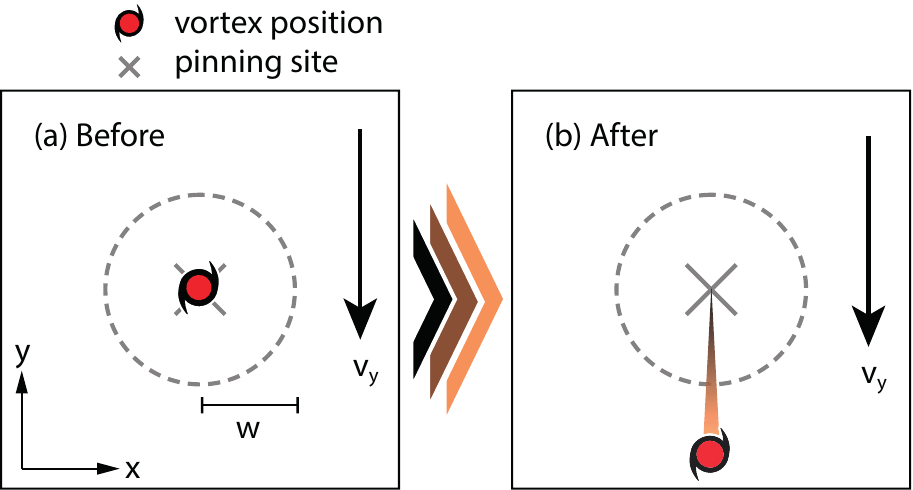}
\caption{Schematic plots of a vortex (red dot) subjected to a pinning site of width $w$ and a background superflow $\mathbf{v}_s=-v_y\hat{\mathbf{y}}$. Initially (a), the vortex is trapped within the pinning site but (b) depins when $v_y$ exceeds a critical value $v_c$. The vortex trajectory is marked by a long triangle with a colour gradient. The color gradient indicates the vortex positions at different times and the width of the triangle is the velocity of the vortex.}
\label{fig:schematic}
\end{center}
\end{figure}

Despite this, the depinning process of a vortex subjected to a background superflow still lacks a detailed theoretical investigation. Thus, we carefully study the depinning dynamics of an initially pinned vortex due to the presence of a background superflow in a two-dimensional superfluid, as illustrated in Fig.~\ref{fig:schematic}. The superfluid system is modelled by the Gross--Pitaevskii equation with a phenomenological dissipation. The phase diagram of the critical depinning velocity is explored in terms of the height and width of the pinning potential, using an energetic argument to understand the transition between the pinned and free vortex states. A complementary set of simulations is also conducted in a different numerical setup with a different boundary condition to validate our finding.

This work is structured as follows. In Sec.~\ref{sec:model} we describe our model and discuss the effects of dissipation on the vortex motion. We then proceed to address the numerical setup of simulating the dynamics of depinning and the results of these simulations in Sec.~\ref{sec:depinning_dynamics}. Subsequently, in Sec.~\ref{sec:critical_velocity}, we propose a phase diagram for the critical depinning velocity and an energy landscape for vortex depinning alongside a validation check of the depinning simulations employing quasiperiodic boundary conditions (QPBCs). Finally, we summarize and discuss our findings in Sec.~\ref{sec:conclusion}.

\section{Theoretical Formalism}\label{sec:model}
\subsection{Gross--Pitaevskii Theory\label{sec:gpe}}
We model a two-dimensional superfluid using the damped Gross--Pitaevskii equation (dGPE)~\cite{Tsubota2002,Penckwitt2002,Tsubota2005,Reeves2013,Reeves2014,Baggaley2018b},
\begin{equation}
i\hbar\frac{\partial}{\partial t}\psi(\mathbf{r},t)=(1-i\gamma)\left(\hat{H}_\mathrm{GP}-\mu\right)\psi(\mathbf{r},t)
\label{eq:dGPE}
\end{equation}
where $\mu$ is the chemical potential and
\begin{equation}
    \hat{H}_\mathrm{GP}=-\frac{\hbar^2\nabla^2}{2m}+V_\mathrm{pin}(\mathbf{r})+g\left|\psi(\mathbf{r},t)\right|^2.
\end{equation}
The parameter $g$ describes the self-repulsion of the superfluid and is related to the $s$-wave interaction strength~\cite{PitaevskiiBook}. For the pinning potential, $V_\mathrm{pin}(\mathbf{r})$, we assume a Gaussian profile~\cite{Sasaki2010,Stagg2014},
\begin{equation}
V_\mathrm{pin}(\mathbf{r})=V_0e^{-(x^2+y^2)/w^2},
\label{eq:pinning_potential}
\end{equation}
with height $V_0$ and width $w$. The dimensionless parameter $\gamma$ in equation~(\ref{eq:dGPE}) represents dissipation arising from the interaction between superfluid and normal components\footnote{Note that we work in the frame of the pinning site and that the normal component is assumed to be at rest in this frame.}.
In the context of Bose--Einstein condensates, $\gamma$ is proportional to temperature and is usually considered to be spatially constant with a numerical value of $\lesssim 10^{-3}$~\cite{Bradley2008, Blakie2008b, Rooney2012}. However, for reasons described in Sec.~\ref{sec:numeric}, our model includes a spatially dependent $\gamma(\mathbf{r})$ that suppresses flows far from the pinning site.

As described later, we choose the value of $\mu$ in all of our simulations such that a constant density, $|\psi|^2=n_0$, is maintained at large distances from the pinning site. The system then has a characteristic length scale given by the healing length, $\xi_0=\hbar/\sqrt{mgn_0}$, and a characteristic time scale $t_0=\hbar/gn_0$. The ratio of these determines the speed of sound in the superfluid, $c_0=\sqrt{gn_0/m}$~\cite{PitaevskiiBook}. Provided that $\gamma \ll 1$, dissipation only plays a role on a much longer time scale, $\tau_0 = t_0/\gamma$~\cite{Bradley2015,Liu2020}.

In this work, we theoretically investigate the critical depinning velocity of a vortex initially pinned by a pinning potential in a superfluid with a background flow $\mathbf{v}_s=-v_y\hat{\mathbf{y}}$.
The width of the pinning potential, $w$, is chosen to be of the same order as the healing length, $\xi_0$, which is the relevant parameter regime for the neutron star crust. Unlike several previous works that assume a wide obstacle, we cannot use either the vortex filament or the Thomas--Fermi~(TF) approximation in this regime. Instead, we use the dGPE model to determine the critical velocity. For later convenience, we introduce the following notation to represent different possible states of the system:
\begin{enumerate}
    \item[(i)] $\psi_{0,v_y}=\sqrt{n_0}e^{-imv_yy/\hbar}$: a homogeneous density solution with a background flow;

    \item[(ii)] $\psi_{\mathrm{v},v_y}$: a single vortex state subjected to a background flow;
    
    \item[(iii)] $\psi_{\mathrm{ps},v_y}$: a vortex-free state with a pinning site subjected to a background flow;
    
    \item[(iv)] $\psi_{\mathrm{pv},v_y}$: a pinned vortex state subjected to a background flow.

    \item[(v)] $\psi_{\mathrm{fv},v_y}$: a free (depinned) vortex state subjected to a background flow far away from the pinning site.
\end{enumerate}
States~(ii)--(v) cannot be fully described analytically and, instead, we study them numerically,
focusing on the dynamics in Sec.~\ref{sec:depinning_dynamics} and the energetics in Sec.~\ref{sec:vortex_energy}.

\subsection{Numerical Setup\label{sec:numeric}}
\begin{figure}
\begin{center}
    \includegraphics[width=1\textwidth]{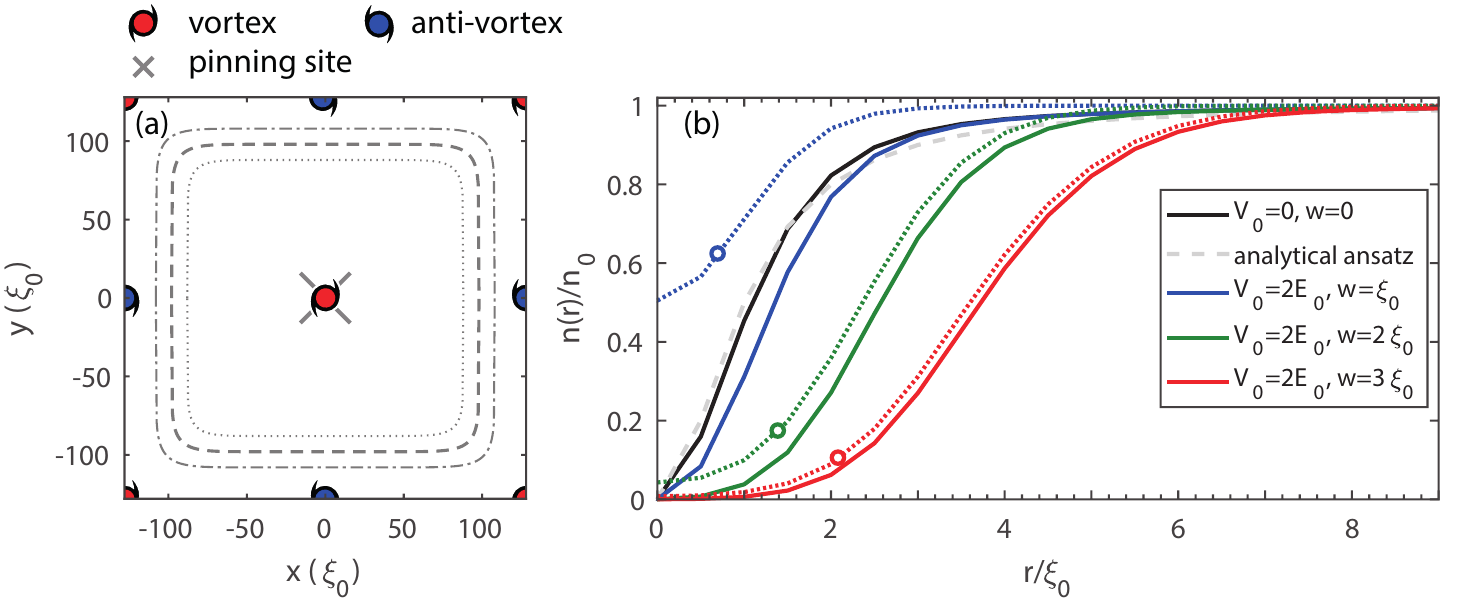}
\caption{(a) A schematic plot of the initialized states with the positions of vortices and anti-vortices marked in red and blue respectively. The targeted vortex sits at the centre while the remaining 3 vortices (1 vortex and 2 anti-vortices) are generated by the phase frustration at the edges due to the periodic boundary condition. The sponge layer Eq.~(\ref{eq:sponge_bc}) is represented by the round rectangular contours for $x^{10}+y^{10}=R_s^{10}$ (grey dashed line), $x^{10}+y^{10}=(R_s - w_\mathrm{abs})^{10}$ (grey dotted line) and $x^{10}+y^{10}=(R_s+w_\mathrm{abs})^{10}$ (grey dashed-dotted line). (b) Density profiles corresponding to the numerical solutions in the absence of a background superflow for a free vortex ($n = |\psi_{\mathrm{v/pv},v_y=0}|^2$, $V_0 = w = 0$) and a pinned vortex ($n = |\psi_{\mathrm{v/pv},v_y=0}|^2$, $V_0=2E_0$ and $w=\lbrace\xi_0,\,2\xi_0,\,3\xi_0\rbrace$. The grey dashed line is the density profile of the analytical ansatz, Eq.~(\ref{eq:vortex_state_appx}), for the free vortex while the dotted lines are the numerical solutions for $\psi_{\mathrm{ps},v_y=0}$. The hollow circles mark the Thomas--Fermi radius $R_\mathrm{TF} = w\ln(V_0/E_0)$ where the density vanishes for $r<R_\mathrm{TF}$ under the TF approximation, suggesting that the vortex density profiles are poorly described by the TF solution.
}
\label{fig:vortex_profile}
\end{center}
\end{figure}
We solve Eq.~(\ref{eq:dGPE}) using a Fourier pseudospectral method and a 4th-order Runge--Kutta scheme with a time step of $\Delta t=10^{-3}t_0$. The computational domain is a square box of width $L_x=L_y=256\xi_0$ with $N_x=N_y=512$ grid points, thereby giving a grid spacing of $\xi_0/2$. The pinning site, when present, is located at the centre of the box.

The use of Fourier transforms requires periodic boundary conditions in both $x$ and $y$, and hence a vanishing net circulation around the boundary of the domain~\cite{rorai2012vortex}. Although we are only interested in the dynamics of a single vortex, it is thus necessary to simulate an alternating lattice of vortices and anti-vortices as illustrated in Fig.~\ref{fig:vortex_profile}~(a). Given the size of the domain, we expect the effect of these additional vortices to be small, and we minimise their effect further by introducing a sponge layer via a spatially dependent dissipation~\cite{Reeves2014,Reeves2015,Rickinson2018}. We adopt a smooth, rounded-rectangular sponge as plotted in Fig.~\ref{fig:vortex_profile}~(a) by defining
\begin{equation}
    \gamma(\mathbf{r})=\gamma_0+\frac{\gamma_\text{abs}}{2}\left[1+\tanh\left\{\frac{\left(x^{10}+y^{10}\right)^{1/10}-R_s}{w_\mathrm{abs}}\right\}\right]
    \label{eq:sponge_bc}
\end{equation}
where $\gamma_0$ is the dissipation induced the interaction between superfluid and normal components. The strong dissipation set by $\gamma_\text{abs}$ at the boundaries of the domain damps any sound waves, thereby isolating the central vortex from its periodic neighbours. However, this dissipation also induces a transverse motion of the vortices near the edge of the domain, with a velocity $-\gamma\kappa v_y\hat{\mathbf{x}}$ \citep{Groszek2018,Bland2022}. Fortunately, this transverse motion ends once the vortices reach the edge of the sponge layer. Thus, for a simulation domain large enough that the sponge layer is sufficiently distant from the central vortex, the effects of the edge vortices are marginal. Here we set $\gamma_\mathrm{abs} = 1$, $R_s=98\xi_0$ and $w_\mathrm{abs}=10\xi_0$.

Another consequence of our periodic boundary conditions is that the imposed superflow velocity, $v_y$, must be an integer multiple of $\Delta v = 2\pi c_0\xi_0/L_y \approx 0.025c_0$. This limits the precision with which the critical velocity can be determined but, given the size of the domain, this does not represent a significant limitation of our methodology.

\subsection{Initialization\label{sec:initializing}}

As noted earlier, we specify the value of the chemical potential $\mu$ in order to fix the (background) density to $n_0$. The correct choice is not only dependent on $g$ and $n_0$ but also on the background superflow velocity $\mathbf{v}_s=-v_y\hat{\mathbf{y}}$. In the dGPE simulation, the dissipative term ultimately drives the system towards a state satisfying $(\hat{H}_\mathrm{GP}-\mu)\psi\approx0$, and so we must choose the value of $\mu$ such that in this steady state the density is equal to $n_0$ far from the pinning site. To determine the appropriate value, we consider the grand-canonical energy of the system,
\begin{equation}
    F=\int_A d^2\mathbf{r}\psi^\ast(\mathbf{r})\left[-\frac{\hbar^2\nabla^2}{2m}+V_\mathrm{pin}(\mathbf{r})+\frac{g}{2}\left|\psi(\mathbf{r})\right|^2\right]\psi(\mathbf{r})-\mu N_A
    \label{eq:free_energy}
\end{equation}
where $A$ is the domain area, $g$ is the $s$-wave interaction strength and $N_A=\int_Ad^2\mathbf{r}\left|\psi(\mathbf{r})\right|^2$ is the particle number in the domain. The damped Gross--Pitaevskii equation~(\ref{eq:dGPE}) can be expressed as $i\hbar\partial_t\psi=(1-i\gamma)\delta F/\delta\psi^\ast$, and the steady state corresponds to a minimum of the the functional $F$. In the absence of any vortex or pinning site, we expect the steady state to be $\psi=\psi_{0,v_y} = \sqrt{n_0}e^{-imv_yy/\hbar}$, and this is a minimum of Eq.~(\ref{eq:free_energy}) provided that
\begin{equation}
    \mu=\mu_{0,v_y}=\frac{mv_y^2}{2}+E_0,
    \label{eq:mu}
\end{equation}
with a background energy $E_0=gn_0$. Therefore, $\mu$ is set to the value given by Eq.~(\ref{eq:mu}) in our simulations.

To initialise the system with a vortex (with or without a pinning site), we first set $v_y=0$ and imprint a phase winding by setting $\psi = \sqrt{n_0}e^{i\phi}$, where $\phi$ is the polar angle relative to the centre of the domain. The GPE is then solved in imaginary time (i.e. with $(1-i\gamma)$ in Eq.~(\ref{eq:dGPE}) replaced by -1), with $\mu = E_0$, until the quantity
\begin{equation}
    \frac{1}{N_A}\int_A d^2\mathbf{r}\,\psi^\ast\hat{H}_\text{GP}\psi
\end{equation}
converges to within $10^{-7}$ of $E_0$. This converged state is then used as the initial condition for the dGPE. As mentioned earlier, because of the periodic boundary conditions this state actually features a lattice of vortices and anti-vortices but, for a sufficiently large domain, the central vortex is essentially independent of the others.

As examples of the solutions of Eq.~\eqref{eq:dGPE}, in Fig.~\ref{fig:vortex_profile}~(b), we plot the density profiles of the free ($\psi_{\mathrm{v},v_y=0}(\mathbf{r})$) and pinned ($\psi_{\mathrm{pv},v_y=0}$) vortices (both represented by solid lines), as well as the vortex-free state subject to the pinning potential ($\psi_{\mathrm{ps},v_y=0}$, represented by dotted lines), for $V_0 = 2E_0$ with different pinning widths $w$. The vortex solutions show a density depletion with a width similar to the pinning site but it is also evident that they do not agree well with the Thomas-Fermi density,
\begin{gather}
    n_\mathrm{TF}(\mathbf{r})=n_0[1-V_\mathrm{pin}(\mathbf{r})/E_0]\Theta(|\mathbf{r}|-R_\mathrm{TF}), \label{eq:thomasfermivortex} \\
    R_\mathrm{TF}=w\sqrt{\ln V_0/E_0}, \label{eq:thomasfermivortexradius}
\end{gather}
as the radii within which the densities are strongly depleted is not in agreement with the TF radii, $\lbrace R_\mathrm{TF}\rbrace$, which we have represented in Fig.~\ref{fig:vortex_profile}~(b) as hollow circles. This suggests that the pinning potential height and width we have specified are too small for a vortex to be in the TF regime~\cite{Stockdale2021}.

\section{Depinning Dynamics\label{sec:depinning_dynamics}}

With the initial state prepared, we impose a background flow by introducing a phase gradient
\begin{equation}
    \psi_{\mathrm{pv},v_y}(\mathbf{r},t=0)=\psi_{\mathrm{pv},v_y=0}(\mathbf{r})e^{-imv_yy/\hbar},
\end{equation}
where the value of $v_y$ is a multiple of $\Delta v=2\pi c_0\xi_0/L_y$ as explained earlier. At the same time we update the value of $\mu$ to $\mu = E_0+mv_y^2/2$ so that the system remains in a steady state far from the vortices and pinning site. The simulations are performed with three values of dissipation: $\gamma_0=0$, $5\times10^{-4}$ and $5\times10^{-3}$. We find that the value of $\gamma_0$ has very little effect on the dynamics, including the depinning process, except that after depinning the vortex drifts with respect to the ambient superflow by an amount proportional to $\gamma_0$. This transverse drift is expected for reasons mentioned in Sec.~\ref{sec:numeric}.

\begin{figure}[tht!]
\includegraphics[width=1\textwidth]{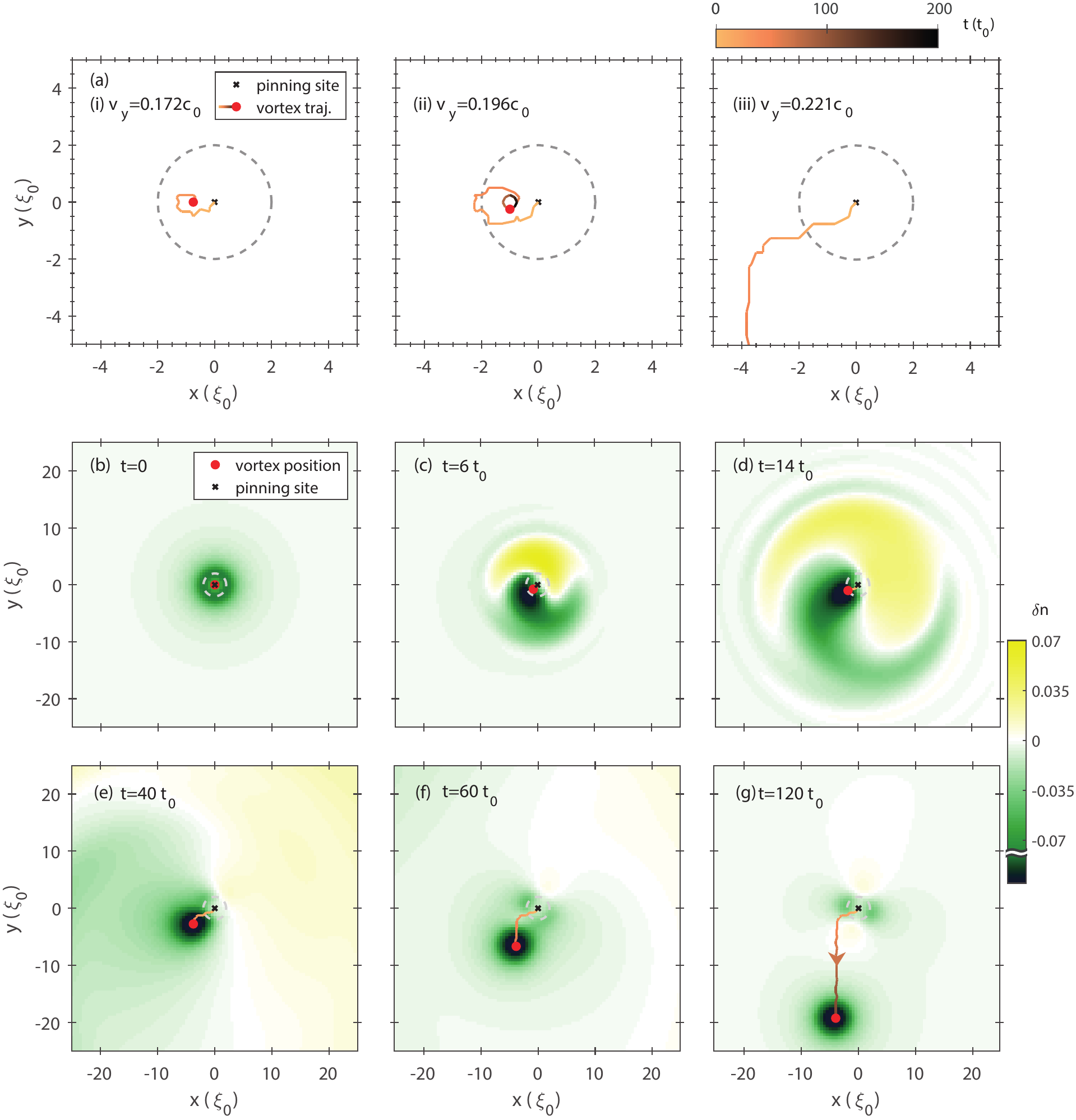}
\caption{(a) Examples of vortex trajectories for $V_0=2E_0$, $w=2\xi_0$ and $\gamma_0=5\times10^{-3}$, with (i) $v_y\approx0.172c_0$, (ii) $v_y\approx0.196c_0$ and (iii) $v_y\approx0.221c_0$ within the time span $t\in(0,200t_0)$. (b) -- (g) The density fluctuations, measured by $\delta n=[n(\mathbf{r},t)-n_\mathrm{ps}(\mathbf{r})]/n_0$, are plotted when (b) $t=0$, (c) $t=6t_0$, (d) $t=14t_0$, (e) $t=40t_0$, (f) $t=60t_0$, (g) $t=120t_0$, where it is evident that the radiated sound waves spiral out from the vortex. The vortex moves to the left of the pinning site due to the Magnus force but, once it escapes from the pinning site, it drifts downward with the superflow at a speed $v_y$.
}\label{fig:unpining_example}
\end{figure}

Three types of dynamics are found in our simulation: (1) if $v_y$ is sufficiently small, then the vortex is displaced a small distance from the centre of the pinning site, but remains pinned; (2) if $v_y$ exceeds a critical value, $v_c$, the vortex depins and is carried away by the ambient superflow; and (3) if $v_y$ is a significant fraction of the sound speed, $c_0$, additional vortices nucleate at the pinning site. However, the present work is not concerned with regime~(3), which has been studied in detail by others, e.g.~Refs.~\citep{Sasaki2010,Stagg2014}. Figure~\ref{fig:unpining_example}~(a) illustrates typical vortex trajectories in regimes~(1) and (2) for $V_0=2E_0$ and $w=2\xi_0$. The vortex position in the Figure is determined by locating the phase defect and density minimum in the wavefunction, after extrapolating to the sub-grid level, and the trajectory is subsequently tracked by linking timeframes using the Hungarian algorithm. In Fig.~\ref{fig:unpining_example} the orange-black colour transition of the trajectories tracks the evolution of the vortex position from $t=0$ to $200t_0$. It is evident that, in regime~(1), the vortex follows a spiral trajectory and eventually reaches an equilibrium position to the left of the pinning site centre. In this new equilibrium, the pinning force $\propto \hat{\mathbf{x}}$ balances the Magnus force $\propto\hat{\mathbf{z}}\times(-v_y\hat{\mathbf{y}})$~\cite{Sonin1997,Groszek2018}. The maximum spiral radius increases with $v_y$, as shown in Fig.~\ref{fig:unpining_example}~(a) (i) and (ii) for $v_y\approx0.172c_0$ and $0.196c_0$ respectively. Conversely, if $v_y >  v_c\approx0.221c_0$ the system is in regime~(2); the vortex initially follows a similar trajectory but moves far enough from the pinning site centre that it escapes and ultimately drifts with the ambient superflow.

In both regimes~(1) and (2) the motion of the vortex excites sound waves that carry energy away from the vortex. To illustrate this, in Fig.~\ref{fig:unpining_example}~(b) -- (g) we plot the density fluctuation $\delta n=[n(\mathbf{r},t)-n_\mathrm{ps}(\mathbf{r})]/n_0$, where $n_\mathrm{ps}(\mathbf{r})=|\psi_{\mathrm{ps},v_y=0}|^2$ is the density in the absence of a vortex or flow, together with the vortex trajectory for $t/t_0=0$, $6$, $14$, $40$, $60$ and $120$. In these plots, the vortex location is marked by the red circle and the orange-black line is the trajectory up to time $t$. Similarly to the case of a vortex in a stirred condensate~\cite{Barenghi2005}, while the vortex remains close to the pinning site the sound waves form a dipolar pattern that spirals out from the vortex. For comparison, we have also performed a simulation with the same parameters but without a vortex; in that regime, sound waves are produced with a circular pattern while the density perturbations are smaller in magnitude by a factor of about 2.

The emission of sound waves becomes negligible in the later time dynamics whether the vortex remains pinned or depins, and those emitted at early times rapidly dissipate within the sponge layer around the domain boundary. We note that, as shown in Fig.~\ref{fig:unpining_example}~(g), density perturbations persist around the pinning site even after the vortex has depinned. This reflects the effect of the ambient flow on the pinning site. Furthermore, as mentioned earlier, at sufficiently high flow velocities vortices are nucleated in the pinning potential~\cite{Sasaki2010,Stagg2014}. If the pinning potential is too high ($V_0\geq3.5E_0$) or too wide ($w\geq3.5\xi_0$) then we find that this regime is reached before the superflow velocity reaches the  critical depinning velocity $v_c$. In this regime the depinning process is complicated by the involvement of multiple vortex interactions~\citep{Stockdale2021} and, therefore, we only present data for pinning sites with smaller values of $V_0$ and $w$.

\section{Critical Depinning Velocity\label{sec:critical_velocity}}

\subsection{Phase Diagram of the Depinning Velocity\label{sec:unpinningphasediagram}}

\begin{figure}
\begin{center}
    \includegraphics[width=1\textwidth]{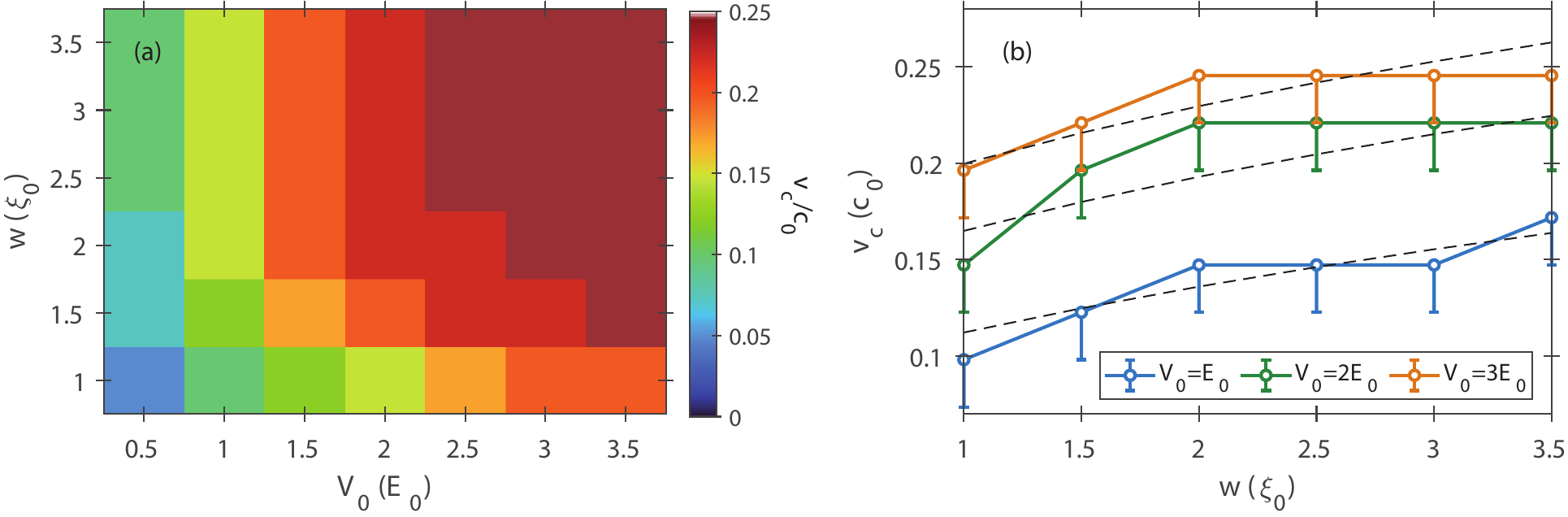}
\caption{(a) Phase diagram of the critical velocity $v_c$ for $\gamma_0=0$ and $5\times10^{-4}$. (b) Comparison between the critical velocity and the empirical formula, Eq.~(\ref{eq:empirical_vc}). The lower errorbar indicates the possible range for $v_c$ given the velocity resolution, $\Delta v\approx0.025c_0$.}
\label{fig:phase_diagram}
\end{center}
\end{figure}

As illustrated in Sec.~\ref{sec:depinning_dynamics}, if the imposed flow, $v_y$, is not sufficient to depin the vortex then it generally settles to a new equilibrium position within a duration of $200t_0$. This new equilibrium is generally within a distance $w$ of the pinning site centre. In our results we therefore consider the vortex to be depinned if it is displaced by more than $1.75w$ within a time of $200t_0$, and we define $v_c$ to be the smallest value of $v_y$ for which depinning occurs. (We recall that our periodic boundary conditions only allow us to increase $v_y$ in steps of $\Delta v\approx0.025c_0$. Therefore our results for $v_c$ may overestimate its true value by up to $\Delta v$.) In Fig.~\ref{fig:phase_diagram}~(a) we show the values of $v_c$ obtained for $V_0/E_0\in[0.5,3.5]$ and $w/\xi_0\in[1,3.5]$. This Figure shows the results in the cases $\gamma_0 = 0$ and $\gamma_0 = 5\times10^{-4}$, which are identical; the results for $\gamma_0 = 5\times10^{-3}$ are nearly identical, except for the case $V_0=0.5E_0$ and $w=\xi_0$, where we find $v_c$ to be larger by $\Delta v$. This indicates that dissipation plays a negligible role in determining the critical velocity in our simulations.

From dimensional analysis, the value of $v_c/c_0$ ought to depend only on $V_0/E_0$, $w/\xi_0$ and $\gamma_0$. Thus, for sufficiently small $\gamma_0$, we ought therefore to be able to fit our results to a function of the form $v_c/c_0 = f(V_0/E_0,w/\xi_0)$. Before performing this fitting, we first briefly review earlier work concerning the form of the function $f$. Stockdale {\it{et al.}}~\cite{Stockdale2021} used the equation of motion of a vortex under the small-displacement approximation~\cite{Groszek2018}, in the absence of dissipation, to investigate the pinning of a vortex in 2D with a circular top-hat barrier and a background flow. Making a Thomas--Fermi approximation for the density, they found that the vortex remains pinned if its equation of motion has a fixed point within the pinning site. In terms of our notation, the critical depinning velocity is then given by
\begin{equation}
    v_c/c_0=\frac{\xi_0}{4\pi}\left(\frac{2w+\delta}{2w\delta}\right)\frac{V_0}{E_0}
    \label{eq:Stockdale_vc}
\end{equation}
where $w$ is the radius of the obstacle and $\delta$ is a phenomenological screening parameter of order $\xi_0$. This result is only valid in the regime where $w > \delta$ and $V_0 < E_0$, however, which is not the regime of interest of the present work.

Earlier, \citet{Schwarz1981} used a vortex-filament model to simulate a vortex pinned to a hemispherical bump on the boundary of the domain. Here the critical depinning velocity was found to be
\begin{equation}
    v_c/c_0\approx\frac{\xi_0}{4\pi D}\ln\frac{w}{\xi_0}
    \label{eq:schwatz}
\end{equation}
where $w$ is the radius of the obstacle and $D$ is the channel width of the superfluid container. Again, this result assumes $w \gg \xi_0$ and is intrinsically three-dimensional, and thus it is not directly applicable to our results.

Motivated by the above literature, and by the results in Fig.~\ref{fig:phase_diagram}~(a), we seek a fit of the critical velocity to the form
\begin{equation}
    v_c/c_0=f\left(\frac{V_0}{E_0},\frac{w}{\xi_0}\right)=a\ln\left[1+ \left(b\frac{w}{\xi_0}+c\right)\frac{V_0}{E_0}\right]
    \label{eq:empirical_vc}
\end{equation}
where the parameters $a$, $b$ and $c$ are assumed to be constant. Applying a least-squares fit to our data, we find
\begin{equation}
    a = 0.1039\pm0.0175, \; b = 0.7581\pm0.4056 \; \mbox{ and } \; c = 1.188\pm0.4880,
\end{equation}
and for these coefficients the empirical formula predicts $v_c$ to within $15\%$ across our data set. In Fig.~\ref{fig:phase_diagram}~(b) we plot $v_c$ as a function of $w$ for three values of $V_0$, alongside our empirical formula.

\subsection{Vortex Energy\label{sec:vortex_energy}}

In our simulations, energy is injected into the system by a sudden increase in the superflow from $0$ to $v_y$. Part of this energy is converted into sound, and part may be used to depin the vortex. In general we would not expect the vortex to depin unless doing so reduces the overall energy in the system, and this fact can potentially be used to determine a lower bound on the critical velocity $v_c$. Specifically, we seek to determine the value of $v_y$ for which the energy of a pinned vortex state exceeds that of an depinned vortex state. This energy can be defined as the energy cost $\alpha_{\psi}$ to create the targeted state from a relevant reference state, which is conventionally chosen to be the ground state of the free energy function given by Eq.~(\ref{eq:free_energy}). For a free vortex state, $\psi_{\mathrm{v,v_y}}$, this is equivalent to the energy of a vortex state, subtracted by the vortex-free state energy, for a fixed number of $N_A$ superfluid atoms in a domain of area $A$~\cite{Fetter1965}. This complexity can be reduced by carrying out the calculation of the grand-canonical energy, Eq.~(\ref{eq:free_energy}), at a fixed chemical potential instead~\cite{PitaevskiiBook,Liu2020a}. In this case,
\begin{equation}
    \alpha_{\text{a},v_y}=F[\psi_\text{a}]-F[\psi_\text{gs}]
\end{equation}
where $\psi_\text{a}$ represents the target state and $\psi_\text{gs}$ is a suitable reference groundstate.
According to the aforementioned energetic argument, a lower bound on $v_c$ can be determined by finding the value of $v_y$ above which the excess energy of a free vortex state far away from the pinning site, $\alpha_{\mathrm{fv},v_y} = \alpha_{\mathrm{v},v_y} + \alpha_{{\mathrm{ps},v_y}}$, is less than that of a pinned vortex, i.e.,
\begin{equation}
\alpha_{\mathrm{fv},v_y}\leq \alpha_{{\mathrm{pv},v_y}}
\label{eq:energy_vc}
\end{equation}
for $v_y\geq v_c$. Here we specifically consider $A$ to be the area of a circular disk with radius $R$, giving $A=\pi R^2$, and a vortex or pinning potential is positioned at the centre of the disk.

\begin{figure}[t!]
\centering
\includegraphics[width=0.6\textwidth]{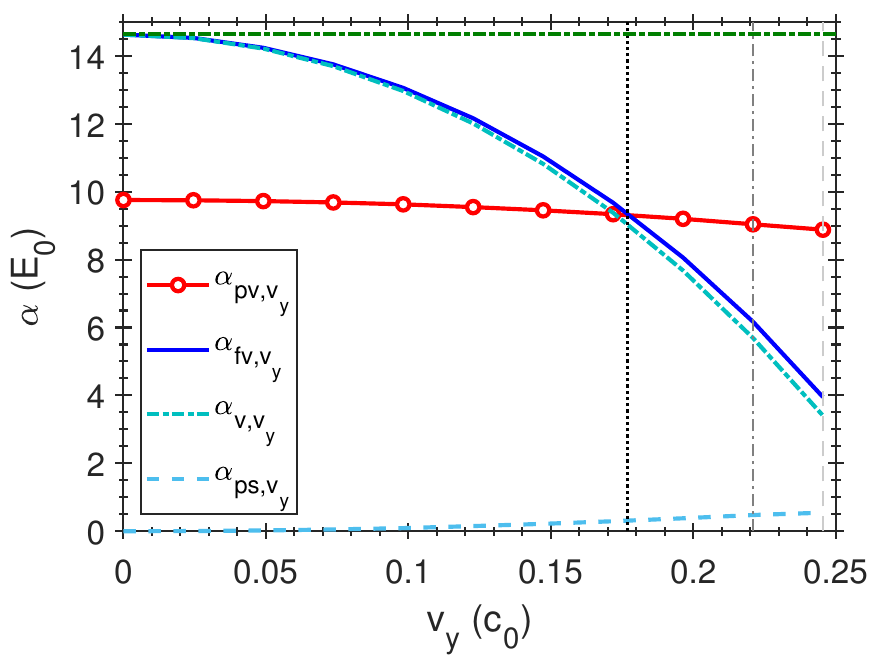}
\caption{The vortex energies for a pinned vortex $\alpha_{\mathrm{pv},v_y}$ and free vortex $\alpha_{\mathrm{fv},v_y}=\alpha_{\mathrm{v},v_y}+\alpha_{\mathrm{ps},v_y}$ as a function of $v_y$ for $V_0=2E_0$ and $w=2\xi_0$. Here, $\alpha_{\mathrm{v},v_y}$ and $\alpha_{\mathrm{ps},v_y}$ are the vortex energies in a homogeneous condensate without a pinning potential and the excess energy to create a flow in a condensate with a present of a barrier, respectively. The green dotted line plots the analytical result of a free vortex, Eq.~(\ref{eq:free_vortex_energy}). The vertical dotted line marks the energetically preferred $v_c$ and the grey dashed-dotted lines are the values of $v_c$ obtained from the dGPE simulation with a background superflow, while the dashed lines are the values of $v_c$ obtained from the complementary, quasiadiabatic advective simulations discussed in Sec.~\ref{sec:quasiperiodicbcs}.}\label{fig:excess_energy}
\end{figure}

The creation energy of a free vortex can be estimated by the required energy added into a homogeneous superfluid, namely, $\alpha_{\mathrm{v},v_y}=F[\psi_{\mathrm{v},v_y}]-F_0$ with $F_0=(mv_y^2-E_0)N_A/2$.
The wavefunction of a free $\kappa$-charged vortex can be well approximated by an analytical profile~\cite{PitaevskiiBook, Fetter1965, Fetter2001},
\begin{equation}
    \psi_{\mathrm{v},v_y}=\sqrt{n_0\frac{r^2/\xi^2}{1+r^2/\xi^2}}e^{i\kappa\theta}e^{-imv_yy/\hbar}
    \label{eq:vortex_state_appx}
\end{equation}
where $r=\sqrt{x^2+y^2}$ and $\xi$ is the radius of the vortex core. In the limit $R\gg\xi$, the leading order terms give
\begin{equation}
    \alpha_{\mathrm{v},v_y}\approx\frac{\pi\hbar^2\kappa^2n_0}{m}\ln\frac{R}{\xi}+\frac{\pi\hbar^2 n_0}{4m}+\frac{\pi gn_0^2\xi^2}{2},
    \label{eq:free_vortex_energy}
\end{equation}
and the minimization of $\alpha_{{\mathrm{v},v_y}}$ with respect to $\xi$ gives $\xi=\left|\kappa\right|\xi_0$~\footnote{From the Laplacian in cylindrical coordinates, $\psi_{\mathrm{v},0}\propto r^{\kappa}$, which may not be fully satisfied by this ansatz for $\kappa>1$.}. This solution, represented in Fig.~\ref{fig:vortex_profile}~(b) as a grey dashed line, agrees well with the corresponding numerical solution represented by the black unbroken line in the same figure. The free vortex energy $\alpha_{\mathrm{v},v_y}$ can then be simplified as\footnote{In Ref.~\cite{PitaevskiiBook,Fetter1965,Fetter2001}, it is shown via the relation $\xi=\hbar/\sqrt{2mgn_0}$ that Eq.~(\ref{eq:free_vortex_energy}) can be rewritten as $\alpha_{\mathrm{v},v_y}\approx\pi\hbar^2\kappa^2m\ln(1.65R/\xi)$},
\begin{equation}
\alpha_{{\mathrm{v},v_y}}\approx\displaystyle\frac{\pi\hbar^2n_0}{m}\ln2.12\frac{R}{\xi_0}
\end{equation}
For a singly-charged vortex, $\kappa=\pm1$, the numerical result of $\psi_{\mathrm{v},v_y=0}$ agrees well with this prediction as demonstrated in Fig.~\ref{fig:excess_energy}. We note, however, that the $v_y$-dependent terms in this analytical calculation cancel each other out and cannot provide a further prediction of $v_c$. Furthermore, we reiterate that due to the healing length-sized pinning potential, the Thomas--Fermi density profile is not a particularly good approximation for the vortex state as shown in Fig.~\ref{fig:vortex_profile}~(b).

\begin{figure}[t!]
\centering
\includegraphics[width=1\textwidth]{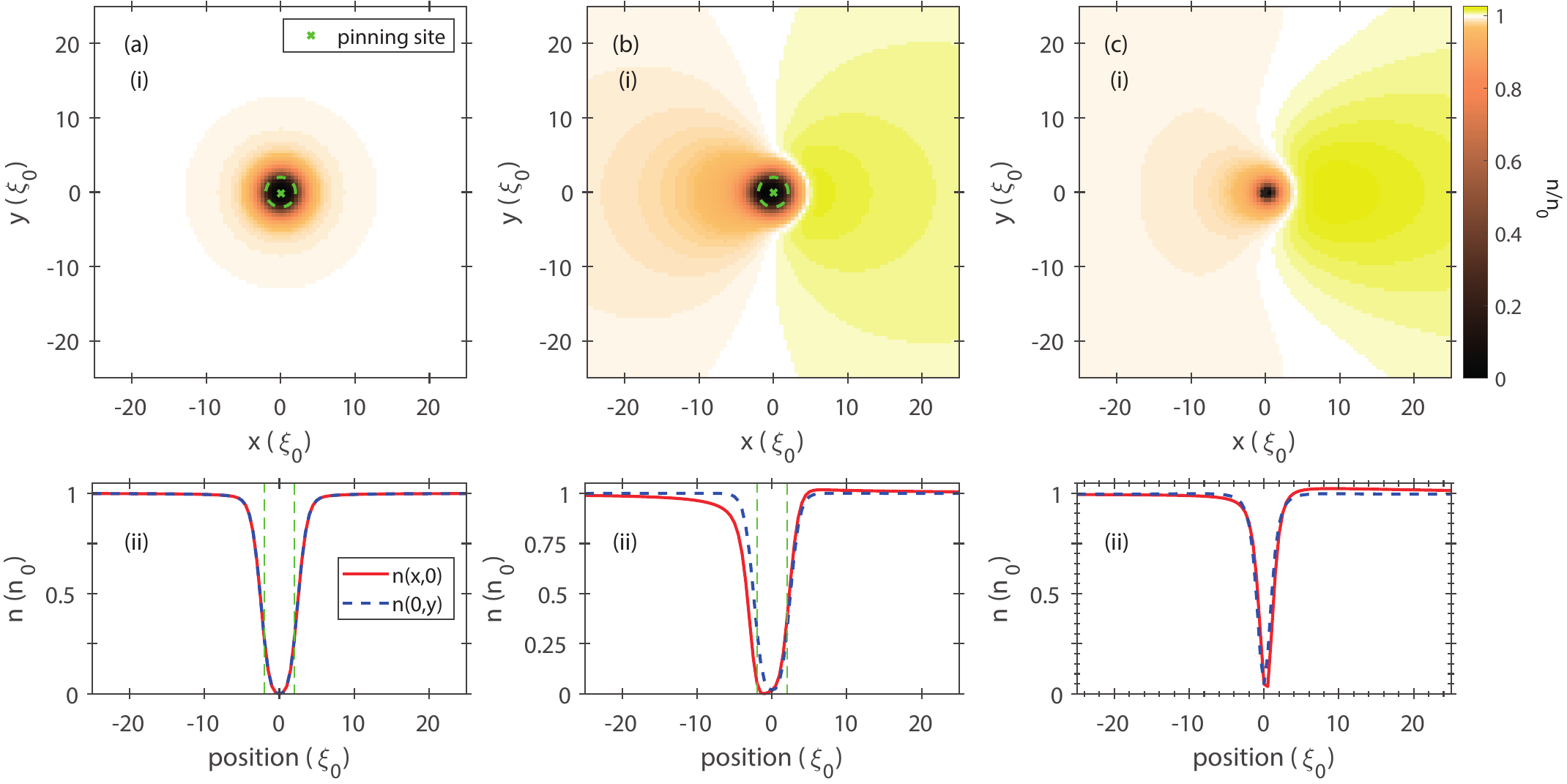}
\caption{Condensate density profiles of $\psi_{\mathrm{pv},v_y}$ for $V_0=2E_0$ and $w=2\xi_0$ with (a) $v_y=0$ and (b) $v_y=0.221c_0$ and (c) that of $\psi_{\mathrm{v},v_y}$ for $v_y=0.221c_0$. Panel (i) presents the two-dimensional profile while panel (ii) shows density slices along the $x$ and $y$ axes with the green dashed lines in (i) and (ii) representing the pinning potential. These clearly depict an axial asymmetry of the density profiles due to the background flow.}
\label{fig:vortex_with_flow}
\end{figure}

To evaluate the vortex energy with a background flow, we numerically obtain $\psi_{\mathrm{pv},v_y}(\mathbf{r})$ and $\psi_{\mathrm{v},v_y}(\mathbf{r})$ by propagating an initial trial vortex state, Eq.~(\ref{eq:vortex_state_appx}), in imaginary time with $\kappa=1$. When the flow amplitude is small, we found that $|\mu_\mathrm{num}-\mu_{0,v_y}|$ can reduce to less than $10^{-7}$ quickly. In contrast for large $v_y$, the convergence is much slower and the central and boundary-induced vortices would annihilate as $\tau\rightarrow\infty$. In addition, when $v_y$ is large enough the vortex is detached from the pinning potential. Hence, we reduce the numerical tolerance so that imaginary time propagation is run until $|\mu_\mathrm{num}-\mu_{0,v_y}|<10^{-6}$. We note, due to this relaxation of numerical tolerance, this state might be merely metastable. Indeed, even for $v_y>v_c$ it is possible to find pinned vortex states and some of the vortices may remain pinned after we move to real-time evolution.

In Fig.~\ref{fig:vortex_with_flow}, we illustrate the density profiles of $\psi_{\mathrm{pv},v_y}$ for $v_y=0$ and $0.221c_0$ with $V_0=2E_0$ and $w=2\xi_0$ as well as that of $\psi_{\mathrm{v},v_y}$ with $v_y\approx0.221c_0$ for the sake of comparison. When the superflow velocity is nonzero, the density becomes locally axially asymmetric about the vortex and pinning site for both $\psi_{\mathrm{pv},v_y}$ and $\psi_{\mathrm{ps},v_y}$. The latter case can be observed in the plot of $\delta n$ in Fig.~\ref{fig:unpining_example}~(g) where a quadrupolar density fluctuation profile is evident~\footnote{We remind the reader that, in our numerical method, $n_0$ is determined by Eq.~(\ref{eq:mu}).}. The combination of the background and vortex flows and the interaction with the pinning potential locally affects the density and has a maximum influence along the direction orthogonal to the flow. Therefore this effect is minimized parallel to the flow in $\psi_{\mathrm{pv},v_y}$, as shown in Fig.~\ref{fig:vortex_with_flow}~(b) and (c), where it is apparent that $n(0,y)$ remains almost symmetric along the $y$ axis. We note that during the propagation of Eq.~\eqref{eq:dGPE} in real time over timescales comparable to the characteristic relaxation timescale, $\hbar/\gamma gn_0=200t_0$, the stability of these vortex states is ambiguous even when $v_y>v_c$. This indicates that a depinning process requires additional energy or instabilities compared to those considered here.

We have also computed $\alpha_\psi$ from the numerical simulations for a domain given by a disk, centred on the pinning site, with an area $A=\pi R^2$ where $R=50\xi_0$. In Fig.~\ref{fig:excess_energy}, we show the numerical results of $\alpha_{{\mathrm{fv},v_y}}$ and $\alpha_{\psi_{\mathrm{pv},v_y}}$ together with $\alpha_{{\mathrm{v},v_y}}$ and $\alpha_{{\mathrm{ps},v_y}}$ for $V_0=2E_0$ and $w=2\xi_0$ as an example. Here, one can see that the pinned vortex energy $\alpha_{{\mathrm{pv},v_y}}$ is approximately constant while the free vortex energy shows a stronger dependence on $v_y$ and decreases monotonically as a function of $v_y$. We also note that the value of $\alpha_{\mathrm{v},v_y=0}$ agrees with the analytical calculation presented earlier. Given that $\alpha_{\mathrm{v},v_y=0} > \alpha_{\mathrm{pv},v_y=0}$ and $d\alpha_{\mathrm{v},v_y}/dv_y > d\alpha_{\mathrm{pv},v_y}/dv_y$, there exists a value of $v_y$ where $\alpha_{\mathrm{fv},v_y} = \alpha_{\mathrm{pv},v_y}$ and Eq.~(\ref{eq:energy_vc}) is satisfied. Above this velocity, we expect it is energetically favourable for the vortex to depin, with the black vertical dotted line in Fig.~\ref{fig:excess_energy} showing the estimated critical velocity for the set of parameters. However, what we observe is that the value of $v_c$ based on energetic considerations alone is found to be less than $v_c$ in the dGPE simulations of Sec.~\ref{sec:depinning_dynamics} by roughly $20\%$ (note the grey vertical dot-dashed line in the figure). Such caveats notwithstanding, we argue that this estimate of $v_c$ based on purely energetic arguments offers a robust order of magnitude estimate. The small difference between the two values points to additional physics playing a role in the process of depinning, which could be the focus of future studies.

\subsection{Quasiadiabatic Advective Acceleration\label{sec:quasiperiodicbcs}}

For reasons explained in Sec.~\ref{sec:numeric}, the use of periodic boundary conditions has consequences for our numerical simulations. For instance, as demonstrated in Fig.~\ref{fig:vortex_profile} (a), the requirement that $\psi$ be circulation-free inside the computational cell results in a vortex lattice of alternating circulations when a vortex is imprinted during the imaginary-time propagation of Eq.~\eqref{eq:dGPE}. Crucially, the periodicity of $\psi$ also restricts the allowed values for the imposed flow velocity $v_y$. To test the robustness of our results, we have therefore also taken a complementary approach, using a different numerical scheme to slowly accelerate the imposed flow from zero to a desired value, after which its value is held fixed. To do so, we first perform a Galilean transformation of the Gross--Pitaevskii equation into the frame moving with velocity $v_y\hat{\mathbf{y}}$, hereafter referred to as the advective Gross--Pitaevskii equation (aGPE):
\begin{equation}
    i\hbar\left(\frac{\partial}{\partial t} - v_y\frac{\partial}{\partial y}\right)\psi(\mathbf{r},t)=\left[-\frac{\hbar^2\nabla^2}{2m}+V_\mathrm{pin}(\mathbf{r})+g\left|\psi(\mathbf{r},t)\right|^2-\mu\right]\psi(\mathbf{r},t). \label{eq:advectivegpe}
\end{equation}
We solve this equation together with quasiperiodic boundary conditions (QPBCs),
\begin{gather}
    \psi(x + L_x, y) = \exp\left[\frac{i\pi}{L_y}\left(y + \frac{L_y}{2}\right)\right]\psi(x, y), \label{eq:qpbcx} \\
    \psi(x, y + L_y) = \psi(x, y), \label{eq:qpbcy}
\end{gather}
that account for the phase winding arising from a single vortex within the domain. Initially introduced in the context of superfluid vortices in Ref.~\cite{Mingarelli2016} and subsequently applied in Refs.~\cite{Wood2019new, Doran2020, Wood2022new}, QPBCs have proven to be advantageous for the study of vortex configurations of nonzero net circulation in a homogeneously uniform background fluid. Because the velocity $v_y$ is slowly increased in this new setup, we anticipate that little sound will be produced, and therefore we do not include any dissipation in the model. It is for this reason that we refer to this setup as ``quasiadiabatic'', in contrast to the instantaneous acceleration of the flow considered in Sec.~\ref{sec:depinning_dynamics}.

\begin{figure}
\centering
\includegraphics[width=0.8\textwidth]{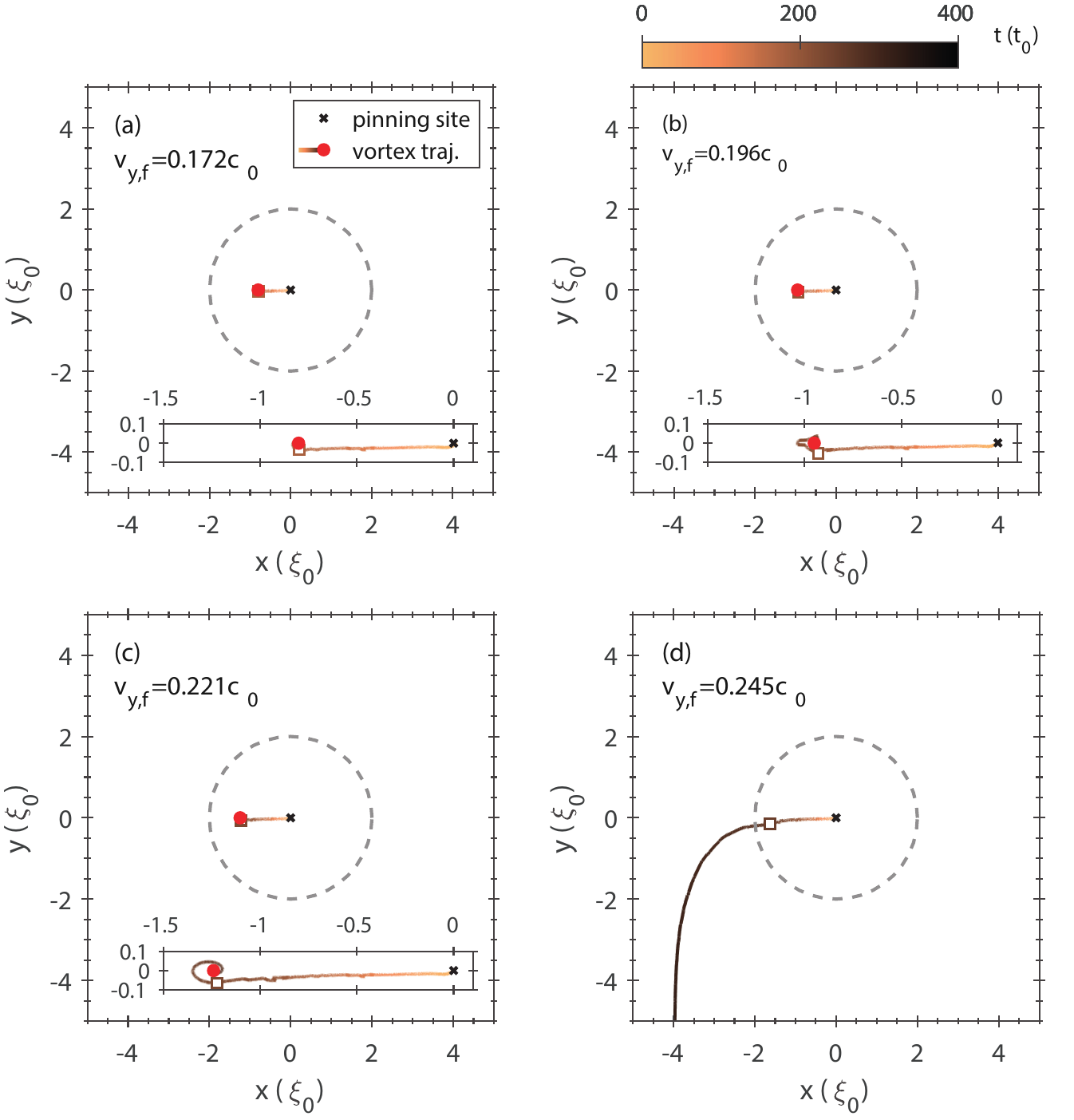}
\caption{Vortex trajectories of the aGPE simulation with a quasiadiabatic acceleration of $v_y(t)$ from zero till a final value (a) $v_{y,f}= 0.174c_0$, (b) $v_{y,f}=0.196c_0$, (c) $v_{y,f}=0.221 c_0$ and (d) $v_{y,f}=0.245c_0$, followed by a duration of evolution at constant $v_y$, for $V_0 = 2E_0$ and $w = 2\xi_0$. The hollow squares mark the vortex position at the end of the advective acceleration and the red circle shows the position of vortex at $t=400t_0$
}
\label{fig:qpbcvortextraj}
\end{figure}

Initially, we obtain the stationary state with a vortex at the origin by imprinting a superfluid phase~\footnote{$\vartheta_3(z; q)$ is the Jacobi theta function of $z$ with nome $q$, whose argument is a suitable phase profile for an incompressible vortex in our chosen (Landau) gauge~\cite{Doran2020}.} \cite{Tkachenko1965, ONeil1989}
\begin{equation}
    \arg\left\lbrace\vartheta_3\left[i\pi\left(\frac{x}{L_y} + \frac{L_x}{L_y}\right) - \pi\left(\frac{y}{L_y} + \frac{1}{2}\right); e^{-\pi\left(\frac{L_y}{L_x}\right)}\right]\right\rbrace \label{eq:jacobitheta}
\end{equation}
and solve the aGPE in imaginary time to find the ground state. Subsequently, the dynamics of this vortex are simulated by propagating Eq.~\eqref{eq:advectivegpe} forward in time with a slow acceleration of the advection as given by
\begin{equation}
    v_y(t)=\left\{\begin{array}{ll}
    \displaystyle\frac{\Delta v}{25t_0}t & t\leq25n_f,
    \\\\
    \displaystyle v_{y,f}=\Delta vn_f, & t > 25n_f,
    \end{array}\right.
    \label{eq:accel}
\end{equation}
where the final advection index $n_f\in\mathbb{Z}$ sets up the advective flow after the acceleration, and the same spatial grids and timestep used are the same as those in Sec.~\ref{sec:unpinningphasediagram}. The advective flow velocity takes $25n_ft_0$ to reach the final value $v_y$ according to Eq.~(\ref{eq:accel}), a duration we believe to be sufficient for the vortex to adjust adiabatically to a finite advection.
Subsequently, the aGPE is propagated up to $t = 400t_0$ to allow time for the vortex to depin, {\`a} la Fig.~\ref{fig:unpining_example}. This numerical setup is designed to inject as little energy as possible into the system during the acceleration of the flow, and therefore we expect the critical depinning velocity, $v_c$, that we obtain to be an upper bound for $v_c$ in any other numerical setup.

Throughout the quasiadiabatic advective simulations, we use a plaquette-based subgrid interpolation method~\cite{Riordan2016} to locate the vortex as precisely as possible and track the evolution of its position. In Fig.~\ref{fig:qpbcvortextraj}, the vortex trajectories of the aGPE simulation for $V_0 = 2E_0$ and $w = 2\xi_0$ with $n_f=7,\;8,\;9,\;10$ are shown. From this, we identify the lowest $v_{y,f}$ that dislodges the initially pinned vortex as the critical velocity. Similar to the simulations presented in Sec.~\ref{sec:depinning_dynamics}, we see that the vortex initially always drifts to the left of the pinning site centre as a consequence of the Magnus effect. Once $v_y$ reaches its final value, $v_{y,f}$, if this value is sufficiently small then the vortex quickly inspirals to a new equilibrium position; this final equilibrium position is comparable to that in Sec.~\ref{sec:depinning_dynamics}. However, if $v_{y,f}$ exceeds a critical velocity, $v_c$, then the vortex is able to escape the pinning potential, as illustrated in Fig.~\ref{fig:qpbcvortextraj}~(d). With this quasiadiabatic setup, we still observe the emission of a spiral pulse of sound waves, but their amplitude is about an order of magnitude smaller than in the instantaneously imposed-flow simulations. The critical velocity obtained via this approach is comparable to that found in Sec.~\ref{sec:depinning_dynamics}, but somewhat larger, as illustrated by the grey dashed vertical line in Fig.~\ref{fig:excess_energy}.

\section{Conclusion\label{sec:conclusion}}

In these proceedings, we have numerically simulated how an imposed superflow can detach a vortex that is initially pinned to an obstacle of size comparable to the superfluid healing length. We find that, below a critical velocity, the Magnus force causes the vortex to drift laterally with respect to the imposed flow before ultimately spiralling into a new equilibrium position. The motion of the vortex is responsible for the emission of a spiral of sound which contributes to the loss of energy and stabilization of the vortex. If the imposed superflow exceeds a critical value, then the vortex escapes the pinning region and is carried away by this superflow. The critical depinning velocities obtained from dGPE simulations are presented in a phase diagram as a function of the height and width of the Gaussian potential which defines the pinning site. Inspired by theoretical predictions existing in the literature~\cite{Stockdale2021,Schwarz1981}, we propose a new empirical formula for the critical depinning velocity which is in quantitative agreement with our numerical findings. When the pinning potential is sufficiently high and wide, vortices can be nucleated in the pinning potential by the imposed flow.

We then provide an energetic argument for a vortex transitioning from a pinned state to a free (depinned) state by evaluating the vortex nucleation energy under different conditions. The free vortex energy can be analytically evaluated by an axially symmetric trial solution, but the flow dependence in energy is not characterized in this Ansatz. Therefore we utilize the imaginary-time propagation method to search for pinned vortex states which are nonexistent when the flow is too rapid. The pinned states are asymmetric, not only in the off-axis location of the vortex in the pinning potential but also a flow-induced asymmetry of the density which is nonetheless small in magnitude. We find that these solutions are not always stationary over long periods of imaginary time propagation but still provide sensible estimations of the pinned states, suggesting that it might be prudent to consider improved numerical methods in future investigations~\cite{Winiecki2000, Liu2020a}. Proceeding to examine the energy landscape of free and pinned vortices we observe that the energy of a vortex decreases monotonically as a function of superflow velocity. However, while the energy of a pinned vortex is lower than that of a free vortex for low velocities, its rate of decrease as a function of the superflow velocity is also correspondingly lower. This results in a crossover superfluid velocity above which free vortices are energetically preferable, a velocity that is comparable to but generally lower than the critical velocities found by direct dGPE simulations with an imposed superflow.

Lastly, we verify our finding with a set of complementary simulations where quasiperiodic boundary conditions are employed to describe a vortex subject to a pinning site being accelerated almost adiabatically to the desired final value of the superflow velocity. These studies, conducted in the reference frame advected with the superflow, provide slightly higher estimates for the critical velocity. Together with our analysis of the energy landscape of pinned and free vortices, this suggests that vortex depinning requires additional energy or instabilities aside from effects arising at the boundaries.

In conclusion, this work develops new insights into the depinning dynamics in a pure 2D superfluid system from a narrow obstacle that are amenable to experimental investigations of superfluids and applicable to theoretical studies of neutron stars. This opens the door to combining studies of the influence of a background superflow with that of ambient sound waves in depinning a vortex from a pinning site. For instance, it has been proposed that the sound waves emitted by a moving vortex around the pinning site can induce the depinning of vortices in its vicinity~\cite{Warszawski2012}. Thus we believe that a natural direction for research in this field would be the study of the vortex-vortex and vortex-sound scattering processes in a system of multiple, initially pinned, vortices in the presence of superflow.


\backmatter





\bmhead{Acknowledgments}

We thank Andrew Groszek and Thomas Bland for their helpful discussions. SBP acknowledges fruitful discussions with Marco Antonelli and Ryan Doran regarding the use of quasiperiodic boundary conditions.

\section*{Declarations}
IKL and SBP conducted the simulations and contributed the majority of the manuscript with a vortex detection algorithm provided by AWB. The concepts of this work were framed by CFB, AWB and TSW with significant physics input, suggestions and guidance regarding the analysis. IKL, AWB, CFB and TSW acknowledge the funding from Science and Technology Facilities Council grant ST/W001020/1, and SBP and AWB acknowledge the support of the Leverhulme Trust through the Research Project Grant RPG-2021-108. The numerical simulations in this investigation were conducted using the Rocket High Performance Computing facility at Newcastle University.

For the purpose of Open Access, the author has applied a CC BY public copyright licence to any Author Accepted Manuscript (AAM) version arising from this submission.


\noindent

\bigskip


\begin{thebibliography}{65}
\ifx \bisbn   \undefined \def \bisbn  #1{ISBN #1}\fi
\ifx \binits  \undefined \def \binits#1{#1}\fi
\ifx \bauthor  \undefined \def \bauthor#1{#1}\fi
\ifx \batitle  \undefined \def \batitle#1{#1}\fi
\ifx \bjtitle  \undefined \def \bjtitle#1{#1}\fi
\ifx \bvolume  \undefined \def \bvolume#1{\textbf{#1}}\fi
\ifx \byear  \undefined \def \byear#1{#1}\fi
\ifx \bissue  \undefined \def \bissue#1{#1}\fi
\ifx \bfpage  \undefined \def \bfpage#1{#1}\fi
\ifx \blpage  \undefined \def \blpage #1{#1}\fi
\ifx \burl  \undefined \def \burl#1{\textsf{#1}}\fi
\ifx \doiurl  \undefined \def \doiurl#1{\url{https://doi.org/#1}}\fi
\ifx \betal  \undefined \def \betal{\textit{et al.}}\fi
\ifx \binstitute  \undefined \def \binstitute#1{#1}\fi
\ifx \binstitutionaled  \undefined \def \binstitutionaled#1{#1}\fi
\ifx \bctitle  \undefined \def \bctitle#1{#1}\fi
\ifx \beditor  \undefined \def \beditor#1{#1}\fi
\ifx \bpublisher  \undefined \def \bpublisher#1{#1}\fi
\ifx \bbtitle  \undefined \def \bbtitle#1{#1}\fi
\ifx \bedition  \undefined \def \bedition#1{#1}\fi
\ifx \bseriesno  \undefined \def \bseriesno#1{#1}\fi
\ifx \blocation  \undefined \def \blocation#1{#1}\fi
\ifx \bsertitle  \undefined \def \bsertitle#1{#1}\fi
\ifx \bsnm \undefined \def \bsnm#1{#1}\fi
\ifx \bsuffix \undefined \def \bsuffix#1{#1}\fi
\ifx \bparticle \undefined \def \bparticle#1{#1}\fi
\ifx \barticle \undefined \def \barticle#1{#1}\fi
\bibcommenthead
\ifx \bconfdate \undefined \def \bconfdate #1{#1}\fi
\ifx \botherref \undefined \def \botherref #1{#1}\fi
\ifx \url \undefined \def \url#1{\textsf{#1}}\fi
\ifx \bchapter \undefined \def \bchapter#1{#1}\fi
\ifx \bbook \undefined \def \bbook#1{#1}\fi
\ifx \bcomment \undefined \def \bcomment#1{#1}\fi
\ifx \oauthor \undefined \def \oauthor#1{#1}\fi
\ifx \citeauthoryear \undefined \def \citeauthoryear#1{#1}\fi
\ifx \endbibitem  \undefined \def \endbibitem {}\fi
\ifx \bconflocation  \undefined \def \bconflocation#1{#1}\fi
\ifx \arxivurl  \undefined \def \arxivurl#1{\textsf{#1}}\fi
\csname PreBibitemsHook\endcsname

\bibitem[\protect\citeauthoryear{Davidenko
  et~al.}{1992}]{davidenko1992stationary}
\begin{barticle}
\bauthor{\bsnm{Davidenko}, \binits{J.M.}},
\bauthor{\bsnm{Pertsov}, \binits{A.V.}},
\bauthor{\bsnm{Salomonsz}, \binits{R.}},
\bauthor{\bsnm{Baxter}, \binits{W.}},
\bauthor{\bsnm{Jalife}, \binits{J.}}:
\batitle{Stationary and drifting spiral waves of excitation in isolated cardiac
  muscle}.
\bjtitle{Nature}
\bvolume{355}(\bissue{6358}),
\bfpage{349}--\blpage{351}
(\byear{1992})
\doiurl{10.1038/355349a0}
\end{barticle}
\endbibitem

\bibitem[\protect\citeauthoryear{Pumir and Krinsky}{1999}]{pumir1999unpinning}
\begin{barticle}
\bauthor{\bsnm{Pumir}, \binits{A.}},
\bauthor{\bsnm{Krinsky}, \binits{V.}}:
\batitle{Unpinning of a rotating wave in cardiac muscle by an electric field}.
\bjtitle{Journal of Theoretical Biology}
\bvolume{199}(\bissue{3}),
\bfpage{311}--\blpage{319}
(\byear{1999})
\doiurl{10.1006/jtbi.1999.0957}
\end{barticle}
\endbibitem

\bibitem[\protect\citeauthoryear{Takagi}{1991}]{Takagi1991}
\begin{barticle}
\bauthor{\bsnm{Takagi}, \binits{S.}}:
\batitle{{Quantum Dynamics and Non-Inertial Frames of References. II: Harmonic
  Oscillators}}.
\bjtitle{Progress of Theoretical Physics}
\bvolume{85}(\bissue{4}),
\bfpage{723}--\blpage{742}
(\byear{1991})
\doiurl{10.1143/ptp/85.4.723}
\end{barticle}
\endbibitem

\bibitem[\protect\citeauthoryear{Paz\'o et~al.}{2004}]{PhysRevLett.93.168303}
\begin{barticle}
\bauthor{\bsnm{Paz\'o}, \binits{D.}},
\bauthor{\bsnm{Kramer}, \binits{L.}},
\bauthor{\bsnm{Pumir}, \binits{A.}},
\bauthor{\bsnm{Kanani}, \binits{S.}},
\bauthor{\bsnm{Efimov}, \binits{I.}},
\bauthor{\bsnm{Krinsky}, \binits{V.}}:
\batitle{Pinning force in active media}.
\bjtitle{Phys. Rev. Lett.}
\bvolume{93},
\bfpage{168303}
(\byear{2004})
\doiurl{10.1103/PhysRevLett.93.168303}
\end{barticle}
\endbibitem

\bibitem[\protect\citeauthoryear{Campbell
  et~al.}{2014}]{PhysRevLett.112.197801}
\begin{barticle}
\bauthor{\bsnm{Campbell}, \binits{M.G.}},
\bauthor{\bsnm{Tasinkevych}, \binits{M.}},
\bauthor{\bsnm{Smalyukh}, \binits{I.I.}}:
\batitle{Topological polymer dispersed liquid crystals with bulk nematic defect
  lines pinned to handlebody surfaces}.
\bjtitle{Phys. Rev. Lett.}
\bvolume{112},
\bfpage{197801}
(\byear{2014})
\doiurl{10.1103/PhysRevLett.112.197801}
\end{barticle}
\endbibitem

\bibitem[\protect\citeauthoryear{Nayek et~al.}{2012}]{nayek2012tailoring}
\begin{barticle}
\bauthor{\bsnm{Nayek}, \binits{P.}},
\bauthor{\bsnm{Jeong}, \binits{H.}},
\bauthor{\bsnm{Park}, \binits{H.R.}},
\bauthor{\bsnm{Kang}, \binits{S.-W.}},
\bauthor{\bsnm{Lee}, \binits{S.H.}},
\bauthor{\bsnm{Park}, \binits{H.S.}},
\bauthor{\bsnm{Lee}, \binits{H.J.}},
\bauthor{\bsnm{Kim}, \binits{H.S.}}:
\batitle{Tailoring monodomain in blue phase liquid crystal by surface pinning
  effect}.
\bjtitle{Applied Physics Express}
\bvolume{5}(\bissue{5}),
\bfpage{051701}
(\byear{2012})
\doiurl{10.1143/APEX.5.051701}
\end{barticle}
\endbibitem

\bibitem[\protect\citeauthoryear{Donnelly}{1991}]{donnelly1991quantized}
\begin{bbook}
\bauthor{\bsnm{Donnelly}, \binits{R.J.}}:
\bbtitle{Quantized Vortices in Helium II}.
\bpublisher{Cambridge University Press}, \blocation{???}
(\byear{1991})
\end{bbook}
\endbibitem

\bibitem[\protect\citeauthoryear{Tsubota and Maekawa}{1993}]{Tsubota1993}
\begin{barticle}
\bauthor{\bsnm{Tsubota}, \binits{M.}},
\bauthor{\bsnm{Maekawa}, \binits{S.}}:
\batitle{{Pinning and depinning of two quantized vortices in superfluid He4}}.
\bjtitle{Physical Review B}
\bvolume{47}(\bissue{18}),
\bfpage{12040}--\blpage{12050}
(\byear{1993})
\doiurl{10.1103/PhysRevB.47.12040}
\end{barticle}
\endbibitem

\bibitem[\protect\citeauthoryear{{Hegde, S. G.; Glaberson}}{1980}]{Hegde1980}
\begin{barticle}
\bauthor{\bsnm{{Hegde, S. G.; Glaberson}}, \binits{W.I.}}:
\batitle{{Pinning of Superfluid Vortices to Surfaces}}.
\bjtitle{Physical Review Letters}
\bvolume{45}(\bissue{3}),
\bfpage{190}--\blpage{193}
(\byear{1980})
\doiurl{10.1103/PhysRevLett.45.190}
\end{barticle}
\endbibitem

\bibitem[\protect\citeauthoryear{Ambegaokar et~al.}{1980}]{PhysRevB.21.1806}
\begin{barticle}
\bauthor{\bsnm{Ambegaokar}, \binits{V.}},
\bauthor{\bsnm{Halperin}, \binits{B.I.}},
\bauthor{\bsnm{Nelson}, \binits{D.R.}},
\bauthor{\bsnm{Siggia}, \binits{E.D.}}:
\batitle{Dynamics of superfluid films}.
\bjtitle{Phys. Rev. B}
\bvolume{21},
\bfpage{1806}--\blpage{1826}
(\byear{1980})
\doiurl{10.1103/PhysRevB.21.1806}
\end{barticle}
\endbibitem

\bibitem[\protect\citeauthoryear{Adams and Glaberson}{1987}]{PhysRevB.35.4633}
\begin{barticle}
\bauthor{\bsnm{Adams}, \binits{P.W.}},
\bauthor{\bsnm{Glaberson}, \binits{W.I.}}:
\batitle{Vortex dynamics in superfluid helium films}.
\bjtitle{Phys. Rev. B}
\bvolume{35},
\bfpage{4633}--\blpage{4652}
(\byear{1987})
\doiurl{10.1103/PhysRevB.35.4633}
\end{barticle}
\endbibitem

\bibitem[\protect\citeauthoryear{Nelson and Vinokur}{1993}]{PhysRevB.48.13060}
\begin{barticle}
\bauthor{\bsnm{Nelson}, \binits{D.R.}},
\bauthor{\bsnm{Vinokur}, \binits{V.M.}}:
\batitle{Boson localization and correlated pinning of superconducting vortex
  arrays}.
\bjtitle{Phys. Rev. B}
\bvolume{48},
\bfpage{13060}--\blpage{13097}
(\byear{1993})
\doiurl{10.1103/PhysRevB.48.13060}
\end{barticle}
\endbibitem

\bibitem[\protect\citeauthoryear{{Blatter, G; Feigel'man, M. V.; Geshkenbein,
  V. B.; Larkin, A. I.; Vinokur}}{1994}]{Blatter1994}
\begin{barticle}
\bauthor{\bsnm{{Blatter, G; Feigel'man, M. V.; Geshkenbein, V. B.; Larkin, A.
  I.; Vinokur}}, \binits{V.M.}}:
\batitle{{Vortices in high-temperature superconductors}}.
\bjtitle{Reviews of Modern Physics}
\bvolume{66}(\bissue{4}),
\bfpage{1125}
(\byear{1994})
\doiurl{10.1103/RevModPhys.66.1125}
\end{barticle}
\endbibitem

\bibitem[\protect\citeauthoryear{Kwok et~al.}{2016}]{kwok2016vortices}
\begin{barticle}
\bauthor{\bsnm{Kwok}, \binits{W.-K.}},
\bauthor{\bsnm{Welp}, \binits{U.}},
\bauthor{\bsnm{Glatz}, \binits{A.}},
\bauthor{\bsnm{Koshelev}, \binits{A.E.}},
\bauthor{\bsnm{Kihlstrom}, \binits{K.J.}},
\bauthor{\bsnm{Crabtree}, \binits{G.W.}}:
\batitle{Vortices in high-performance high-temperature superconductors}.
\bjtitle{Reports on Progress in Physics}
\bvolume{79}(\bissue{11}),
\bfpage{116501}
(\byear{2016})
\doiurl{10.1088/0034-4885/79/11/116501}
\end{barticle}
\endbibitem

\bibitem[\protect\citeauthoryear{Neely et~al.}{2013}]{PhysRevLett.111.235301}
\begin{barticle}
\bauthor{\bsnm{Neely}, \binits{T.W.}},
\bauthor{\bsnm{Bradley}, \binits{A.S.}},
\bauthor{\bsnm{Samson}, \binits{E.C.}},
\bauthor{\bsnm{Rooney}, \binits{S.J.}},
\bauthor{\bsnm{Wright}, \binits{E.M.}},
\bauthor{\bsnm{Law}, \binits{K.J.H.}},
\bauthor{\bsnm{Carretero-Gonz\'alez}, \binits{R.}},
\bauthor{\bsnm{Kevrekidis}, \binits{P.G.}},
\bauthor{\bsnm{Davis}, \binits{M.J.}},
\bauthor{\bsnm{Anderson}, \binits{B.P.}}:
\batitle{Characteristics of two-dimensional quantum turbulence in a
  compressible superfluid}.
\bjtitle{Phys. Rev. Lett.}
\bvolume{111},
\bfpage{235301}
(\byear{2013})
\doiurl{10.1103/PhysRevLett.111.235301}
\end{barticle}
\endbibitem

\bibitem[\protect\citeauthoryear{Amico et~al.}{2021}]{Amico:2020eej}
\begin{barticle}
\bauthor{\bsnm{Amico}, \binits{L.}}, \betal:
\batitle{{Roadmap on Atomtronics: State of the art and perspective}}.
\bjtitle{AVS Quantum Sci.}
\bvolume{3}(\bissue{3}),
\bfpage{039201}
(\byear{2021})
\doiurl{10.1116/5.0026178}
{\href{https://arxiv.org/abs/2008.04439}{{arXiv:2008.04439}}}
{[cond-mat.quant-gas]}
\end{barticle}
\endbibitem

\bibitem[\protect\citeauthoryear{Bland et~al.}{2022}]{Bland2022}
\begin{barticle}
\bauthor{\bsnm{Bland}, \binits{T.}},
\bauthor{\bsnm{Yatsuta}, \binits{I.V.}},
\bauthor{\bsnm{Edwards}, \binits{M.}},
\bauthor{\bsnm{Nikolaieva}, \binits{Y.O.}},
\bauthor{\bsnm{Oliinyk}, \binits{A.O.}},
\bauthor{\bsnm{Yakimenko}, \binits{A.I.}},
\bauthor{\bsnm{Proukakis}, \binits{N.P.}}:
\batitle{{Persistent current oscillations in a double-ring quantum gas}}.
\bjtitle{Physical Review Research}
\bvolume{4}(\bissue{4}),
\bfpage{1}--\blpage{11}
(\byear{2022})
\doiurl{10.1103/PhysRevResearch.4.043171}
{\href{https://arxiv.org/abs/2204.14120}{{arXiv:2204.14120}}}
\end{barticle}
\endbibitem

\bibitem[\protect\citeauthoryear{Schive et~al.}{2014}]{Schive2014}
\begin{barticle}
\bauthor{\bsnm{Schive}, \binits{H.-Y.}},
\bauthor{\bsnm{Chiueh}, \binits{T.}},
\bauthor{\bsnm{Broadhurst}, \binits{T.}}:
\batitle{{Cosmic structure as the quantum interference of a coherent dark
  wave}}.
\bjtitle{Nature Physics}
\bvolume{10}(\bissue{7}),
\bfpage{496}--\blpage{499}
(\byear{2014})
\doiurl{10.1038/nphys2996}
\end{barticle}
\endbibitem

\bibitem[\protect\citeauthoryear{Marsh and Pop}{2015}]{Marsh2015}
\begin{barticle}
\bauthor{\bsnm{Marsh}, \binits{D.J.E.}},
\bauthor{\bsnm{Pop}, \binits{A.-R.}}:
\batitle{{Axion dark matter, solitons and the cusp--core problem}}.
\bjtitle{Monthly Notices of the Royal Astronomical Society}
\bvolume{451}(\bissue{3}),
\bfpage{2479}--\blpage{2492}
(\byear{2015})
\doiurl{10.1093/mnras/stv1050}
\end{barticle}
\endbibitem

\bibitem[\protect\citeauthoryear{Marsh}{2016}]{Marsh2016}
\begin{barticle}
\bauthor{\bsnm{Marsh}, \binits{D.J.E.}}:
\batitle{{Axion cosmology}}.
\bjtitle{Physics Reports}
\bvolume{643},
\bfpage{1}--\blpage{79}
(\byear{2016})
\doiurl{10.1016/j.physrep.2016.06.005}
{\href{https://arxiv.org/abs/1510.07633}{{arXiv:1510.07633}}}
\end{barticle}
\endbibitem

\bibitem[\protect\citeauthoryear{Ferreira}{2021}]{Ferreira}
\begin{barticle}
\bauthor{\bsnm{Ferreira}, \binits{E.G.M.}}:
\batitle{{Ultra-Light Dark Matter}}.
\bjtitle{The Astronomy and Astrophysics Review}
\bvolume{29},
\bfpage{7}
(\byear{2021})
\doiurl{10.1007/s00159-021-00135-6}
\end{barticle}
\endbibitem

\bibitem[\protect\citeauthoryear{Rindler-Daller and
  Shapiro}{2012}]{Rindler-Daller2012}
\begin{barticle}
\bauthor{\bsnm{Rindler-Daller}, \binits{T.}},
\bauthor{\bsnm{Shapiro}, \binits{P.R.}}:
\batitle{{Angular momentum and vortex formation in Bose-Einstein-condensed cold
  dark matter haloes}}.
\bjtitle{Monthly Notices of the Royal Astronomical Society}
\bvolume{422}(\bissue{1}),
\bfpage{135}--\blpage{161}
(\byear{2012})
\doiurl{10.1111/j.1365-2966.2012.20588.x}
\end{barticle}
\endbibitem

\bibitem[\protect\citeauthoryear{Dmitriev et~al.}{2021}]{Dmitriev2021}
\begin{barticle}
\bauthor{\bsnm{Dmitriev}, \binits{A.S.}},
\bauthor{\bsnm{Levkov}, \binits{D.G.}},
\bauthor{\bsnm{Panin}, \binits{A.G.}},
\bauthor{\bsnm{Pushnaya}, \binits{E.K.}},
\bauthor{\bsnm{Tkachev}, \binits{I.I.}}:
\batitle{{Instability of rotating Bose stars}}.
\bjtitle{Physical Review D}
\bvolume{104}(\bissue{2}),
\bfpage{023504}
(\byear{2021})
\doiurl{10.1103/PhysRevD.104.023504}
{\href{https://arxiv.org/abs/2104.00962}{{arXiv:2104.00962}}}
\end{barticle}
\endbibitem

\bibitem[\protect\citeauthoryear{Schobesberger
  et~al.}{2021}]{Schobesberger2021}
\begin{barticle}
\bauthor{\bsnm{Schobesberger}, \binits{S.O.}},
\bauthor{\bsnm{Rindler-Daller}, \binits{T.}},
\bauthor{\bsnm{Shapiro}, \binits{P.R.}}:
\batitle{{Angular Momentum and the Absence of Vortices in the Cores of Fuzzy
  Dark Matter Haloes}}.
\bjtitle{Monthly Notices of the Royal Astronomical Society}
\bvolume{505}(\bissue{1}),
\bfpage{802}--\blpage{829}
(\byear{2021})
\doiurl{10.1093/mnras/stab1153}
{\href{https://arxiv.org/abs/2101.04958}{{arXiv:2101.04958}}}
\end{barticle}
\endbibitem

\bibitem[\protect\citeauthoryear{Liu et~al.}{2023}]{Liu2023}
\begin{barticle}
\bauthor{\bsnm{Liu}, \binits{I.-K.}},
\bauthor{\bsnm{Proukakis}, \binits{N.P.}},
\bauthor{\bsnm{Rigopoulos}, \binits{G.}}:
\batitle{{Coherent and incoherent structures in fuzzy dark matter haloes}}.
\bjtitle{Monthly Notices of the Royal Astronomical Society}
\bvolume{521}(\bissue{3}),
\bfpage{3625}--\blpage{3647}
(\byear{2023})
\doiurl{10.1093/mnras/stad591}
{\href{https://arxiv.org/abs/2211.02565}{{arXiv:2211.02565}}}
\end{barticle}
\endbibitem

\bibitem[\protect\citeauthoryear{Pethick et~al.}{2017}]{Pethick2017}
\begin{botherref}
\oauthor{\bsnm{Pethick}, \binits{C.J.}},
\oauthor{\bsnm{Sch{\"{a}}fer}, \binits{T.}},
\oauthor{\bsnm{Schwenk}, \binits{A.}}:
{Bose-Einstein condensates in neutron stars}.
Universal Themes of Bose-Einstein Condensation,
573--592
(2017)
\doiurl{10.1017/9781316084366.031}
{\href{https://arxiv.org/abs/1507.05839}{{arXiv:1507.05839}}}
\end{botherref}
\endbibitem

\bibitem[\protect\citeauthoryear{Jones}{2001}]{Jones2001}
\begin{barticle}
\bauthor{\bsnm{Jones}, \binits{P.B.}}:
\batitle{{First-principles point-defect calculations for solid neutron star
  matter}}.
\bjtitle{Monthly Notices of the Royal Astronomical Society}
\bvolume{175},
\bfpage{167}--\blpage{175}
(\byear{2001})
\doiurl{10.1046/j.1365-8711.2001.03990.x}
\end{barticle}
\endbibitem

\bibitem[\protect\citeauthoryear{Donati and Pizzochero}{2004}]{Donati2004}
\begin{barticle}
\bauthor{\bsnm{Donati}, \binits{P.}},
\bauthor{\bsnm{Pizzochero}, \binits{P.M.}}:
\batitle{{Fully consistent semi-classical treatment of vortex – nucleus
  interaction in rotating neutron stars}}.
\bjtitle{Nuclear Physics A}
\bvolume{742},
\bfpage{363}--\blpage{379}
(\byear{2004})
\doiurl{10.1016/j.nuclphysa.2004.07.002}
\end{barticle}
\endbibitem

\bibitem[\protect\citeauthoryear{Drummond and Melatos}{2017}]{Drummond2017a}
\begin{barticle}
\bauthor{\bsnm{Drummond}, \binits{L.V.}},
\bauthor{\bsnm{Melatos}, \binits{A.}}:
\batitle{{Stability of interlinked neutron vortex and proton flux tube arrays
  in a neutron star: equilibrium configurations}}.
\bjtitle{Monthly Notices of the Royal Astronomical Society}
\bvolume{472}(\bissue{4}),
\bfpage{4851}--\blpage{4869}
(\byear{2017})
\doiurl{10.1093/mnras/stx2301}
\end{barticle}
\endbibitem

\bibitem[\protect\citeauthoryear{Drummond and Melatos}{2018}]{Drummond2018}
\begin{barticle}
\bauthor{\bsnm{Drummond}, \binits{L.V.}},
\bauthor{\bsnm{Melatos}, \binits{A.}}:
\batitle{{Stability of interlinked neutron vortex and proton flux-tube arrays
  in a neutron star -- II. Far-from-equilibrium dynamics}}.
\bjtitle{Monthly Notices of the Royal Astronomical Society}
\bvolume{475}(\bissue{1}),
\bfpage{910}--\blpage{920}
(\byear{2018})
\doiurl{10.1093/mnras/stx3197}
\end{barticle}
\endbibitem

\bibitem[\protect\citeauthoryear{L{\"{o}}nnborn et~al.}{2019}]{Lonnborn2019}
\begin{barticle}
\bauthor{\bsnm{L{\"{o}}nnborn}, \binits{J.R.}},
\bauthor{\bsnm{Melatos}, \binits{A.}},
\bauthor{\bsnm{Haskell}, \binits{B.}}:
\batitle{{Collective, glitch-like vortex motion in a neutron star with an
  annular pinning barrier}}.
\bjtitle{Monthly Notices of the Royal Astronomical Society}
\bvolume{487}(\bissue{1}),
\bfpage{702}--\blpage{710}
(\byear{2019})
\doiurl{10.1093/mnras/stz1302}
{\href{https://arxiv.org/abs/1905.02877}{{arXiv:1905.02877}}}
\end{barticle}
\endbibitem

\bibitem[\protect\citeauthoryear{Sonin}{1997}]{Sonin1997}
\begin{barticle}
\bauthor{\bsnm{Sonin}, \binits{E.}}:
\batitle{{Magnus force in superfluids and superconductors}}.
\bjtitle{Physical Review B}
\bvolume{55}(\bissue{1}),
\bfpage{485}--\blpage{501}
(\byear{1997})
\doiurl{10.1103/PhysRevB.55.485}
{\href{https://arxiv.org/abs/9606099}{{arXiv:9606099}}}
{[cond-mat]}
\end{barticle}
\endbibitem

\bibitem[\protect\citeauthoryear{Schwarz}{1981}]{Schwarz1981}
\begin{barticle}
\bauthor{\bsnm{Schwarz}, \binits{K.W.}}:
\batitle{{Vortex pinning in superfluid helium}}.
\bjtitle{Physical Review Letters}
\bvolume{47}(\bissue{4}),
\bfpage{251}--\blpage{254}
(\byear{1981})
\doiurl{10.1103/PhysRevLett.47.251}
\end{barticle}
\endbibitem

\bibitem[\protect\citeauthoryear{Stockdale et~al.}{2021}]{Stockdale2021}
\begin{barticle}
\bauthor{\bsnm{Stockdale}, \binits{O.R.}},
\bauthor{\bsnm{Reeves}, \binits{M.T.}},
\bauthor{\bsnm{Davis}, \binits{M.J.}}:
\batitle{{Dynamical Mechanisms of Vortex Pinning in Superfluid Thin Films}}.
\bjtitle{Physical Review Letters}
\bvolume{127}(\bissue{25}),
\bfpage{255302}
(\byear{2021})
\doiurl{10.1103/PhysRevLett.127.255302}
{\href{https://arxiv.org/abs/2102.04712}{{arXiv:2102.04712}}}
\end{barticle}
\endbibitem

\bibitem[\protect\citeauthoryear{Groszek et~al.}{2018}]{Groszek2018}
\begin{barticle}
\bauthor{\bsnm{Groszek}, \binits{A.J.}},
\bauthor{\bsnm{Paganin}, \binits{D.M.}},
\bauthor{\bsnm{Helmerson}, \binits{K.}},
\bauthor{\bsnm{Simula}, \binits{T.P.}}:
\batitle{{Motion of vortices in inhomogeneous Bose-Einstein condensates}}.
\bjtitle{Physical Review A}
\bvolume{97}(\bissue{2}),
\bfpage{023617}
(\byear{2018})
{\href{https://arxiv.org/abs/1708.09202}{{1708.09202}}}
\end{barticle}
\endbibitem

\bibitem[\protect\citeauthoryear{Tsubota et~al.}{2002}]{Tsubota2002}
\begin{barticle}
\bauthor{\bsnm{Tsubota}, \binits{M.}},
\bauthor{\bsnm{Kasamatsu}, \binits{K.}},
\bauthor{\bsnm{Ueda}, \binits{M.}}:
\batitle{{Vortex lattice formation in a rotating Bose-Einstein condensate}}.
\bjtitle{Physical Review A}
\bvolume{65}(\bissue{2}),
\bfpage{023603}
(\byear{2002})
\doiurl{10.1103/PhysRevA.65.023603}
{\href{https://arxiv.org/abs/0104523}{{arXiv:0104523}}}
{[cond-mat]}
\end{barticle}
\endbibitem

\bibitem[\protect\citeauthoryear{Penckwitt and Ballagh}{2002}]{Penckwitt2002}
\begin{barticle}
\bauthor{\bsnm{Penckwitt}, \binits{A.A.}},
\bauthor{\bsnm{Ballagh}, \binits{R.J.}}:
\batitle{{Nucleation, Growth, and Stabilization of Bose-Einstein Condensate
  Vortex Lattices}}.
\bjtitle{Physical Review Letters}
\bvolume{89}(\bissue{22}),
\bfpage{260402}
(\byear{2002})
\doiurl{10.1103/PhysRevLett.89.260402}
\end{barticle}
\endbibitem

\bibitem[\protect\citeauthoryear{Tsubota et~al.}{2005}]{Tsubota2005}
\begin{barticle}
\bauthor{\bsnm{Tsubota}, \binits{M.}},
\bauthor{\bsnm{Kasamatsu}, \binits{K.}},
\bauthor{\bsnm{Ueda}, \binits{M.}},
\bauthor{\bsnm{Machida}, \binits{M.}},
\bauthor{\bsnm{Sasa}, \binits{N.}},
\bauthor{\bsnm{Tsubota}, \binits{M.}}:
\batitle{{Three-dimensional dynamics of vortex-lattice formation in
  Bose-Einstein condensates}}.
\bjtitle{Physical Review A}
\bvolume{71}(\bissue{6}),
\bfpage{063616}
(\byear{2005})
\doiurl{10.1103/PhysRevA.71.063616}
{\href{https://arxiv.org/abs/0104523}{{arXiv:0104523}}}
{[cond-mat]}
\end{barticle}
\endbibitem

\bibitem[\protect\citeauthoryear{Reeves et~al.}{2013}]{Reeves2013}
\begin{barticle}
\bauthor{\bsnm{Reeves}, \binits{M.T.}},
\bauthor{\bsnm{Billam}, \binits{T.P.}},
\bauthor{\bsnm{Anderson}, \binits{B.P.}},
\bauthor{\bsnm{Bradley}, \binits{A.S.}}:
\batitle{{Inverse energy cascade in forced two-dimensional quantum
  turbulence}}.
\bjtitle{Physical Review Letters}
\bvolume{110}(\bissue{10}),
\bfpage{104501}
(\byear{2013})
\doiurl{10.1103/PhysRevLett.110.104501}
\end{barticle}
\endbibitem

\bibitem[\protect\citeauthoryear{Reeves et~al.}{2014}]{Reeves2014}
\begin{barticle}
\bauthor{\bsnm{Reeves}, \binits{M.T.}},
\bauthor{\bsnm{Billam}, \binits{T.P.}},
\bauthor{\bsnm{Anderson}, \binits{B.P.}},
\bauthor{\bsnm{Bradley}, \binits{A.S.}}:
\batitle{{Signatures of coherent vortex structures in a disordered
  two-dimensional quantum fluid}}.
\bjtitle{Physical Review A}
\bvolume{89}(\bissue{5}),
\bfpage{053631}
(\byear{2014})
\doiurl{10.1103/PhysRevA.89.053631}
\end{barticle}
\endbibitem

\bibitem[\protect\citeauthoryear{Baggaley and Barenghi}{2018}]{Baggaley2018b}
\begin{barticle}
\bauthor{\bsnm{Baggaley}, \binits{A.W.}},
\bauthor{\bsnm{Barenghi}, \binits{C.F.}}:
\batitle{Decay of homogeneous two-dimensional quantum turbulence}.
\bjtitle{Phys. Rev. A}
\bvolume{97},
\bfpage{033601}
(\byear{2018})
\doiurl{10.1103/PhysRevA.97.033601}
\end{barticle}
\endbibitem

\bibitem[\protect\citeauthoryear{Pitaevskii and
  Stringari}{2016}]{PitaevskiiBook}
\begin{bbook}
\bauthor{\bsnm{Pitaevskii}, \binits{L.}},
\bauthor{\bsnm{Stringari}, \binits{S.}}:
\bbtitle{Bose-Einstein Condensation and Superfluidity}
vol. \bseriesno{164}.
\bpublisher{Oxford University Press}, \blocation{???}
(\byear{2016}).
\doiurl{10.1093/acprof:oso/9780198758884.001.0001} .
\burl{https://doi.org/10.1093/acprof:oso/9780198758884.001.0001}
\end{bbook}
\endbibitem

\bibitem[\protect\citeauthoryear{Sasaki et~al.}{2010}]{Sasaki2010}
\begin{barticle}
\bauthor{\bsnm{Sasaki}, \binits{K.}},
\bauthor{\bsnm{Suzuki}, \binits{N.}},
\bauthor{\bsnm{Saito}, \binits{H.}}:
\batitle{{B{\'{e}}nard-von K{\'{a}}rm{\'{a}}n vortex street in a bose-einstein
  condensate}}.
\bjtitle{Physical Review Letters}
\bvolume{104}(\bissue{15}),
\bfpage{150404}
(\byear{2010})
\doiurl{10.1103/PhysRevLett.104.150404}
\end{barticle}
\endbibitem

\bibitem[\protect\citeauthoryear{Stagg et~al.}{2014}]{Stagg2014}
\begin{barticle}
\bauthor{\bsnm{Stagg}, \binits{G.W.}},
\bauthor{\bsnm{Parker}, \binits{N.G.}},
\bauthor{\bsnm{Barenghi}, \binits{C.F.}}:
\batitle{{Quantum analogues of classical wakes in Bose–Einstein
  condensates}}.
\bjtitle{Journal of Physics B: Atomic, Molecular and Optical Physics}
\bvolume{47}(\bissue{9}),
\bfpage{095304}
(\byear{2014})
\doiurl{10.1088/0953-4075/47/9/095304}
\end{barticle}
\endbibitem

\bibitem[\protect\citeauthoryear{Bradley et~al.}{2008}]{Bradley2008}
\begin{botherref}
\oauthor{\bsnm{Bradley}, \binits{A.S.}},
\oauthor{\bsnm{Gardiner}, \binits{C.W.}},
\oauthor{\bsnm{Davis}, \binits{M.J.}}:
{Bose-Einstein condensation from a rotating thermal cloud: Vortex nucleation
  and lattice formation}
\textbf{77}(3),
033616
(2008)
\doiurl{10.1103/PhysRevA.77.033616}
\end{botherref}
\endbibitem

\bibitem[\protect\citeauthoryear{Blakie et~al.}{2008}]{Blakie2008b}
\begin{botherref}
\oauthor{\bsnm{Blakie}, \binits{P.B.}},
\oauthor{\bsnm{Bradley}, \binits{A.S.}},
\oauthor{\bsnm{Davis}, \binits{M.J.}},
\oauthor{\bsnm{Ballagh}, \binits{R.J.}},
\oauthor{\bsnm{Gardiner}, \binits{C.W.}}:
{Dynamics and statistical mechanics of ultra-cold Bose gases using c-field
  techniques}
\textbf{57}(5),
363--455
(2008)
\doiurl{10.1080/00018730802564254}
\end{botherref}
\endbibitem

\bibitem[\protect\citeauthoryear{Rooney et~al.}{2012}]{Rooney2012}
\begin{barticle}
\bauthor{\bsnm{Rooney}, \binits{S.J.}},
\bauthor{\bsnm{Blakie}, \binits{P.B.}},
\bauthor{\bsnm{Bradley}, \binits{A.S.}}:
\batitle{{Stochastic projected Gross-Pitaevskii equation}}.
\bjtitle{Physical Review A}
\bvolume{86}(\bissue{5}),
\bfpage{053634}
(\byear{2012})
\doiurl{10.1103/PhysRevA.86.053634}
\end{barticle}
\endbibitem

\bibitem[\protect\citeauthoryear{Bradley et~al.}{2015}]{Bradley2015}
\begin{barticle}
\bauthor{\bsnm{Bradley}, \binits{A.S.}},
\bauthor{\bsnm{Rooney}, \binits{S.J.}},
\bauthor{\bsnm{McDonald}, \binits{R.G.}}:
\batitle{{Low-dimensional stochastic projected Gross-Pitaevskii equation}}.
\bjtitle{Physical Review A}
\bvolume{92}(\bissue{3}),
\bfpage{033631}
(\byear{2015})
\doiurl{10.1103/PhysRevA.92.033631}
\end{barticle}
\endbibitem

\bibitem[\protect\citeauthoryear{Liu et~al.}{2020}]{Liu2020}
\begin{barticle}
\bauthor{\bsnm{Liu}, \binits{I.-K.}},
\bauthor{\bsnm{Dziarmaga}, \binits{J.}},
\bauthor{\bsnm{Gou}, \binits{S.-C.}},
\bauthor{\bsnm{Dalfovo}, \binits{F.}},
\bauthor{\bsnm{Proukakis}, \binits{N.P.}}:
\batitle{{Kibble-Zurek dynamics in a trapped ultracold Bose gas}}.
\bjtitle{Physical Review Research}
\bvolume{2}(\bissue{3}),
\bfpage{033183}
(\byear{2020})
\doiurl{10.1103/PhysRevResearch.2.033183}
\end{barticle}
\endbibitem

\bibitem[\protect\citeauthoryear{Rorai}{2012}]{rorai2012vortex}
\begin{botherref}
\oauthor{\bsnm{Rorai}, \binits{C.}}:
Vortex reconnection in superfluid helium.
PhD thesis,
Universit{\`a} degli studi di Trieste
(2012)
\end{botherref}
\endbibitem

\bibitem[\protect\citeauthoryear{Reeves et~al.}{2015}]{Reeves2015}
\begin{barticle}
\bauthor{\bsnm{Reeves}, \binits{M.T.}},
\bauthor{\bsnm{Billam}, \binits{T.P.}},
\bauthor{\bsnm{Anderson}, \binits{B.P.}},
\bauthor{\bsnm{Bradley}, \binits{A.S.}}:
\batitle{{Identifying a Superfluid Reynolds Number via Dynamical Similarity}}.
\bjtitle{Physical Review Letters}
\bvolume{114}(\bissue{15}),
\bfpage{155302}
(\byear{2015})
\doiurl{10.1103/PhysRevLett.114.155302}
\end{barticle}
\endbibitem

\bibitem[\protect\citeauthoryear{Rickinson et~al.}{2018}]{Rickinson2018}
\begin{barticle}
\bauthor{\bsnm{Rickinson}, \binits{E.}},
\bauthor{\bsnm{Parker}, \binits{N.G.}},
\bauthor{\bsnm{Baggaley}, \binits{A.W.}},
\bauthor{\bsnm{Barenghi}, \binits{C.F.}}:
\batitle{{Diffusion of quantum vortices}}.
\bjtitle{Physical Review A}
\bvolume{98}(\bissue{2}),
\bfpage{023608}
(\byear{2018})
\doiurl{10.1103/PhysRevA.98.023608}
{\href{https://arxiv.org/abs/1805.09187}{{arXiv:1805.09187}}}
\end{barticle}
\endbibitem

\bibitem[\protect\citeauthoryear{Barenghi et~al.}{2005}]{Barenghi2005}
\begin{barticle}
\bauthor{\bsnm{Barenghi}, \binits{C.F.}},
\bauthor{\bsnm{Parker}, \binits{N.G.}},
\bauthor{\bsnm{Proukakis}, \binits{N.P.}},
\bauthor{\bsnm{Adams}, \binits{C.S.}}:
\batitle{{Decay of quantised vorticity by sound emission}}.
\bjtitle{Journal of Low Temperature Physics}
\bvolume{138}(\bissue{3-4}),
\bfpage{629}--\blpage{634}
(\byear{2005})
\doiurl{10.1007/s10909-005-2272-5}
{\href{https://arxiv.org/abs/0405635}{{arXiv:0405635}}}
{[cond-mat]}
\end{barticle}
\endbibitem

\bibitem[\protect\citeauthoryear{Fetter}{1965}]{Fetter1965}
\begin{barticle}
\bauthor{\bsnm{Fetter}, \binits{A.L.}}:
\batitle{{Vortices in an Imperfect Bose Gas. I. The Condensate}}.
\bjtitle{Physical Review}
\bvolume{138}(\bissue{2A}),
\bfpage{429}
(\byear{1965})
\end{barticle}
\endbibitem

\bibitem[\protect\citeauthoryear{Liu et~al.}{2020}]{Liu2020a}
\begin{barticle}
\bauthor{\bsnm{Liu}, \binits{I.-K.}},
\bauthor{\bsnm{Gou}, \binits{S.-C.}},
\bauthor{\bsnm{Takeuchi}, \binits{H.}}:
\batitle{{Phase Diagram of Solitons in the Polar Phase of a Spin-1
  Bose-Einstein Condensate}}.
\bjtitle{Physical Review Research}
\bvolume{2}(\bissue{3}),
\bfpage{33506}
(\byear{2020})
\doiurl{10.1103/PhysRevResearch.2.033506}
{\href{https://arxiv.org/abs/2002.06088}{{arXiv:2002.06088}}}
\end{barticle}
\endbibitem

\bibitem[\protect\citeauthoryear{Fetter and Svidzinsky}{2001}]{Fetter2001}
\begin{botherref}
\oauthor{\bsnm{Fetter}, \binits{A.L.}},
\oauthor{\bsnm{Svidzinsky}, \binits{A.A.}}:
{Vortices in a trapped dilute Bose-Einstein condensate}.
Journal of Physics Condensed Matter
\textbf{13}(12)
(2001)
\doiurl{10.1088/0953-8984/13/12/201}
{\href{https://arxiv.org/abs/0102003}{{arXiv:0102003}}}
{[cond-mat]}
\end{botherref}
\endbibitem

\bibitem[\protect\citeauthoryear{Mingarelli et~al.}{2016}]{Mingarelli2016}
\begin{barticle}
\bauthor{\bsnm{Mingarelli}, \binits{L.}},
\bauthor{\bsnm{Keaveny}, \binits{E.E.}},
\bauthor{\bsnm{Barnett}, \binits{R.}}:
\batitle{{Simulating infnite vortex lattices in superfluids}}.
\bjtitle{Journal of Physics: Condensed Matter}
\bvolume{28}(\bissue{28}),
\bfpage{285201}
(\byear{2016})
\doiurl{10.1088/0953-8984/28/28/285201}
\end{barticle}
\endbibitem

\bibitem[\protect\citeauthoryear{Wood et~al.}{2019}]{Wood2019new}
\begin{barticle}
\bauthor{\bsnm{Wood}, \binits{T.S.}},
\bauthor{\bsnm{Mesgarnezhad}, \binits{M.}},
\bauthor{\bsnm{Stagg}, \binits{G.W.}},
\bauthor{\bsnm{Barenghi}, \binits{C.F.}}:
\batitle{{Quasiperiodic boundary conditions for three-dimensional
  superfluids}}.
\bjtitle{Physical Review B: Condensed Matter and Materials Physics}
\bvolume{100}(\bissue{2}),
\bfpage{020405}
(\byear{2019})
\doiurl{10.1103/PhysRevB.100.024505}
\end{barticle}
\endbibitem

\bibitem[\protect\citeauthoryear{Doran and Billam}{2020}]{Doran2020}
\begin{barticle}
\bauthor{\bsnm{Doran}, \binits{R.}},
\bauthor{\bsnm{Billam}, \binits{T.P.}}:
\batitle{{Numerical method for the projected Gross-Pitaevskii equation in an
  infinite rotating two-dimensional Bose gas}}.
\bjtitle{Physical Review E}
\bvolume{102}(\bissue{3}),
\bfpage{033309}
(\byear{2020})
\doiurl{10.1103/PhysRevE.102.033309}
\end{barticle}
\endbibitem

\bibitem[\protect\citeauthoryear{Wood and Graber}{2022}]{Wood2022new}
\begin{barticle}
\bauthor{\bsnm{Wood}, \binits{T.S.}},
\bauthor{\bsnm{Graber}, \binits{V.}}:
\batitle{{Superconducting Phases in Neutron Star Cores}}.
\bjtitle{Journal of Physics: Condensed Matter}
\bvolume{8}(\bissue{4}),
\bfpage{228}
(\byear{2022})
\doiurl{10.3390/universe8040228}
\end{barticle}
\endbibitem

\bibitem[\protect\citeauthoryear{Tkachenko}{1965}]{Tkachenko1965}
\begin{barticle}
\bauthor{\bsnm{Tkachenko}, \binits{V.K.}}:
\batitle{{On Vortex Lattices}}.
\bjtitle{Journal of Experimental and Theoretical Physics (USSR)}
\bvolume{49}(\bissue{6}),
\bfpage{1875}--\blpage{1883}
(\byear{1965})
\end{barticle}
\endbibitem

\bibitem[\protect\citeauthoryear{O'Neil}{1989}]{ONeil1989}
\begin{barticle}
\bauthor{\bsnm{O'Neil}, \binits{K.A.}}:
\batitle{{On the Hamiltonian dynamics of vortex lattices }}.
\bjtitle{Journal of Mathematical Physics}
\bvolume{30}(\bissue{6}),
\bfpage{1373}--\blpage{1379}
(\byear{1989})
\doiurl{10.1063/1.528605}
\end{barticle}
\endbibitem

\bibitem[\protect\citeauthoryear{Riordan et~al.}{2016}]{Riordan2016}
\begin{botherref}
\oauthor{\bsnm{Riordan}, \binits{L.J.O.}},
\oauthor{\bsnm{Busch}, \binits{T.}},
\oauthor{\bsnm{Riordan}, \binits{L.E.E.J.O.}},
\oauthor{\bsnm{Busch}, \binits{T.}}:
{Topological defect dynamics of vortex lattices in Bose-Einstein condensates}
\textbf{94}(5),
053603
(2016)
\doiurl{10.1103/PhysRevA.94.053603}
\end{botherref}
\endbibitem

\bibitem[\protect\citeauthoryear{Winiecki and Adams}{2000}]{Winiecki2000}
\begin{barticle}
\bauthor{\bsnm{Winiecki}, \binits{T.}},
\bauthor{\bsnm{Adams}, \binits{C.S.}}:
\batitle{{Motion of an object through a quantum fluid}}.
\bjtitle{Europhysics Letters}
\bvolume{52}(\bissue{3}),
\bfpage{257}--\blpage{263}
(\byear{2000})
\doiurl{10.1209/epl/i2000-00432-x}
\end{barticle}
\endbibitem

\bibitem[\protect\citeauthoryear{Warszawski et~al.}{2012}]{Warszawski2012}
\begin{barticle}
\bauthor{\bsnm{Warszawski}, \binits{L.}},
\bauthor{\bsnm{Melatos}, \binits{A.}},
\bauthor{\bsnm{Berloff}, \binits{N.G.}}:
\batitle{{Unpinning triggers for superfluid vortex avalanches}}.
\bjtitle{Physical Review B}
\bvolume{85}(\bissue{10}),
\bfpage{104503}
(\byear{2012})
\doiurl{10.1103/PhysRevB.85.104503}
\end{barticle}
\endbibitem

\end{thebibliography}

\end{document}